%% file: main_arxiv_v2.tex
\documentclass[11pt]{article}
\pdfoutput=1
\usepackage{jheparxiv}
\usepackage{mathletters}

\usepackage[dvipsnames]{xcolor}

\usepackage{bm}
\usepackage{amsmath}
\usepackage{amssymb}
\usepackage{graphicx}
\usepackage{hyperref}
\usepackage{slashed}
\usepackage{amsfonts}
\usepackage{amsbsy}
\usepackage{color}
\usepackage{array}
\usepackage{dcolumn}
\usepackage{tensor}
\usepackage{braket}
\usepackage{mathrsfs}
\usepackage{tikz}
\usepackage{verbatim}

\newcommand{\cE}{\mathcal{E}}
\newcommand{\e}{\mathrm{e}}
\newcommand{\im}{\mathrm{i}}
\newcommand{\cN}{\mathcal{N}}
\newcommand{\cM}{\mathcal{M}}
\newcommand{\cA}{\mathcal{A}}
\newcommand{\la}{\langle}
\newcommand{\ra}{\rangle}

\begin{document}

\title{One-loop multicollinear limits from $2$-point amplitudes on self-dual backgrounds}

\author[a]{Tim Adamo,}
\emailAdd{t.adamo@ed.ac.uk}

\author[b]{Anton Ilderton}
\emailAdd{anton.ilderton@ed.ac.uk}

\author[c]{\& Alexander J.~MacLeod}
\emailAdd{alexander.macleod@eli-beams.eu}

\affiliation[a]{School of Mathematics and Maxwell Institute for Mathematical Sciences \\
        University of Edinburgh, EH9 3FD, United Kingdom}

\affiliation[b]{Higgs Centre, School of Physics and Astronomy \\
 University of Edinburgh, EH9 3FD, United Kingdom}

\affiliation[c]{Institute of Physics of the ASCR, ELI-Beamlines \\
 Na Slovance 2, 18221 Prague, Czech Republic}

\abstract{For scattering amplitudes in strong background fields, it is -- at least in principle -- possible to perturbatively expand the background to obtain higher-point vacuum amplitudes. In the case of self-dual plane wave backgrounds we consider this expansion for two-point, one-loop amplitudes in pure Yang-Mills, QED and QCD. This enables us to obtain multicollinear limits of 1-loop vacuum amplitudes; the resulting helicity configurations are surprisingly restricted, with only the all-plus helicity amplitude surviving. These results are shown to be consistent with well-known vacuum amplitudes. We also show that for both abelian and non-abelian supersymmetric gauge theories, there is no helicity flip (and hence no vacuum birefringence) on any plane wave background, generalising a result previously known in the Euler-Heisenberg limit of super-QED.}

\maketitle

\section{Introduction \label{sec:Intro}}
\input{intro.tex}

\section{QED \label{sec:QED}}
\input{QED.tex}

\section{Yang-Mills \label{sec:YM}}
\input{YangMills.tex}

\section{$\mathcal{N}=1$ SUSY gauge theories\label{sec:SUSY}}
\input{SUSY.tex}


\section{Conclusions\label{sec:CONCS}}

\input{Conclusions.tex}

\acknowledgments
We thank Lance Dixon for many useful discussions and suggestions, and Tom Heinzl for comments on a draft of this manuscript.  TA is supported by a Royal Society University Research Fellowship; AI and AJM are supported by the EPSRC, grant EP EP/S010319/1.


\appendix

\section{All-plus amplitude in massive QED \label{app:MASS}}
\input{massive.tex}

\section{Consistency check: non-chiral backgrounds and higher-points \label{app:six-point}}
\input{appendix-six-point.tex}

\section{Yang-Mills helicity non-flip on a general plane wave background
\label{app:YMNonFlip}}

\input{Appendix_No_Flip_YM.tex}

\bibliographystyle{JHEP}
\bibliography{Bib}

\end{document}

%% file: intro.tex

Modern techniques in quantum field theory (QFT) have given access to all-multiplicity expressions for scattering amplitudes at tree-level and beyond (cf., \cite{Elvang:2013cua,Dixon:2013uaa,Cheung:2017pzi}). 
While most progress has been made for amplitudes in vacuum, coupling QFT to non-trivial backgrounds offers an environment in which to explore novel physical effects and their underlying structures.

Amplitudes in weak backgrounds can be calculated perturbatively in powers of an appropriate constant parametrizing the strength of the background. 
Strong backgrounds, however, typically require the introduction of non-perturbative methods. 
An appropriate framework is background field perturbation theory~\cite{Furry:1951zz,DeWitt:1967ub,tHooft:1975uxh,Boulware:1980av,Abbott:1981ke}, in which the strong background is treated without approximation, while scattering of particles on the background is treated in perturbation theory. 
The practical use of this method relies, though, on being able to solve the classical dynamics of particles in the chosen background exactly (or at least without recourse to perturbation theory).

When this is possible, the resulting amplitudes can be far more complex than those in vacuum, and exhibit new structures with important phenomenological consequences. 
For instance, processes which are kinematically forbidden in vacuum have non-trivial amplitudes in strong backgrounds. 
In strong field quantum electrodynamics (QED), where scattering amplitudes in strong backgrounds have been studied for decades~\cite{Nikishov:1964zza,Ritus:1985,DiPiazza:2011tq,Seipt:2017ckc,King:2015tba}, there are many interesting examples, including $1\to2$ non-linear Compton scattering and $1\to3$ trident pair-production, with both being of past and future experimental interest~\cite{Bamber:1999zt,Abramowicz:2021zja,E320}. 
A particularly famous example is that of vacuum birefringence~\cite{Toll:1952rq,Heinzl:2006xc}; while usually studied in low-energy (constant background) Euler-Heisenberg effective theory~\cite{Dunne:2004nc,Gies2017}, it is underpinned by the one-loop $1\to1$ photon helicity flip amplitude on a background~\cite{Dinu:2013gaa}.

Given such background field amplitudes, one can -- perhaps counter-intuitively -- re-expand in powers of the background and, stripping away the background itself, gain access to vacuum amplitudes. 
These will be restricted in momentum/helicity configuration by the support of the initial background configuration but, notably, may be of arbitrarily high multiplicity in external particle number. 
Exploring this expansion can provide insight into differences and similarities between amplitudes in vacuum and in backgrounds, as well as useful theoretical `data'; examples include extracting data for the gauge theory side of an explicit double copy to gravity beyond flat backgrounds (e.g., \cite{Adamo:2017nia,Farrow:2018yni,Adamo:2018mpq,Lipstein:2019mpu,Adamo:2020qru,Albayrak:2020fyp,Armstrong:2020woi}), or for alternative bases of external states (e.g., \cite{Casali:2020vuy,Casali:2020uvr,Pasterski:2020pdk}).

In this paper we consider scattering on \emph{self-dual} plane-wave backgrounds in QED, Yang-Mills and QCD. 
The motivation for this choice is three-fold. 
First, plane waves provide, due to their high degree of symmetry, an analytically tractable setup for the calculation of amplitudes on strong backgrounds. 
Second, amplitudes on self-dual backgrounds can exhibit a remarkable and unexpected simplicity~\cite{Dunne:2002qf,Dunne:2002qg}; indeed there are all-multiplicity formulae for tree-level scattering amplitudes of Yang-Mills theory on these backgrounds~\cite{Adamo:2020syc,Adamo:2020yzi}. 
Third, self-duality of the background cleanly controls the helicity configurations of vacuum amplitudes generated by perturbative expansion, since every photon or gluon `extracted' from the background will be of positive helicity (with all-incoming conventions).  

We focus on 2-point, one-loop scattering amplitudes on the background, and the extraction from them of $N$-point, one-loop amplitudes in vacuum, for arbitrary $N$. 
On the self-dual background, the (non-vanishing) 2-point amplitudes have all-incoming helicity configurations $({+}{+})$ or $({-}{+})$, which correspond to the helicity flip and non-flip amplitudes, respectively, in `physical' in/out conventions. Consequently, the $N$-point, one-loop vacuum amplitudes resulting from perturbative expansion of the background will be in the helicity configurations $(\pm +\cdots +)$. 
These configurations are interesting for many reasons: they are rational functions when the electron/quark mass is zero, include the leading-colour contribution to 1-loop gluon scattering, and have all-multiplicity formulae (when the electron/quark is massless)~\cite{Mahlon:1993fe,Bern:1993qk,Mahlon:1993si,Bern:1994ju,Bern:2005ji}. Remarkably, the all-plus helicity configuration in massless QCD is related to maximally supersymmetric Yang-Mills amplitudes with two negative helicity gluons~\cite{Bern:1996ja,Britto:2020crg}, and this configuration is even linked with ultraviolet divergences in two-loop perturbative gravity~\cite{Bern:2015xsa,Bern:2017puu,Chattopadhyay:2020oxe}.

We find that, due to the properties of plane waves, the perturbative expansion of the background-field amplitude reproduces the maximally mulicollinear (non-trivial) limits of vacuum amplitudes. 
The limit and identification of recovered vacuum amplitudes can be quite subtle, as
we explore in both the massless and massive cases.

In the presence of supersymmetry, vacuum amplitudes for the $({\pm}{+}\cdots{+})$ helicity configurations  vanish to all loop orders as a consequence of supersymmetric Ward identities~\cite{Grisaru:1976vm,Grisaru:1977px,Parke:1985pn,Kunszt:1985mg}. 
This vanishing should be reflected in the perturbative expansion of the helicity flip and non-flip amplitudes for super-QED and super-Yang-Mills on a self-dual plane wave background. 
In fact, we find a \emph{much} stronger result: helicity flip amplitudes of super-QED and super-Yang-Mills vanish on general plane wave backgrounds of \emph{any} shape and polarization.
This is an all-energies generalization of a result which was previously known only in the Euler-Heisenberg regime of super-QED, and implies that there can be no vacuum birefringence in supersymmetric gauge theories.

\medskip 

This paper is organised as follows. 
In the remainder of this section we introduce plane wave backgrounds and illustrate the essential structures which are encountered throughout by reviewing two-point, one-loop photon amplitudes in QED. 
We also explain the relation to vacuum amplitudes through perturbative expansions in the background. 
In Section~\ref{sec:QED} we perform this expansion in massless and massive QED to obtain highly collinear limits of all-plus, one-minus-rest-plus and split helicity vacuum amplitudes, and provide consistency checks with the literature. 
We consider analogous helicity amplitudes for gluon scattering on Yang-Mills plane wave backgrounds in Section~\ref{sec:YM} for pure Yang-Mills theory and QCD.  
In Section~\ref{sec:SUSY} we extend our results beyond self-dual backgrounds, showing that helicity flip cannot occur on any plane wave background in supersymmetric theories.
Section~\ref{sec:CONCS} concludes, while appendices~\ref{app:MASS} --~\ref{app:YMNonFlip} include technical details of various calculations and consistency checks.


\subsection{Two-point QED amplitudes on a plane wave background}
We work in lightfront coordinates on $d$-dimensional Minkowski space,
\be\label{metric}
	\ud s^2 = 2\ud x^\LCp \ud x^\LCm -\ud x^\LCperp \ud x^\LCperp \;,
\ee
in which $\perp = 1,\ldots d-2$ labels the `transverse' directions and $x^\LCm$ is `lightfront time'. 
An electromagnetic plane wave is a highly symmetric solution of Maxwell's equations in vacuum, and in lightfront coordinates is described by the potential\footnote{In other applications, particularly for comparison with results for gravitational plane waves, it is convenient to use the gauge equivalent potential $eA_\text{bg}(x) = - x^\LCperp a'_\LCperp(x^\LCm)\ud x^\LCm$. 
Mapped to the gravitational setting, this corresponds to the choice of using Brinkmann coordinates~\cite{Brinkmann:1925fr}, which are global; see~\cite{Adamo:2017nia,Ilderton:2018lsf} for discussions.}
\be\label{eA}
	eA_\text{bg}(x)  = a_\LCperp(x^\LCm)\, \ud x^\LCperp \;,
\ee
with $a_\LCpm =0$. We assume that the components $a_\perp(x^-)$ obey the `sandwich' condition -- that they are compactly supported in lightfront time $x^\LCm$ -- but are otherwise freely chosen. 
Plane waves posses $2d-3$ symmetries, the generators of which form a Heisenberg algebra with centre $\partial_\LCp$, generating a covariantly constant null symmetry. 
In terms of scattering amplitudes, these symmetries imply, as is also clear from the form of \eqref{eA}, that $d-1$ momentum components $k_\LCp, k_\LCperp$ remain conserved; as such any $N$-point QED amplitude takes the form  
\be\label{A-delta-M}
	(2\pi)^{d-1}\delta^{d-1}_{\LCp,\LCperp}( k_1 + k_2+ \cdots +k_N)\,\mathcal{M}\;,
\ee
where the $k_{j\,\mu}$ are (all-incoming) particle momenta, and $\cM$ is the non-trivial part of the amplitude, on which we focus.  The sandwich condition on the plane wave background
ensures that these scattering amplitudes are well-defined~\cite{Schwinger:1951nm,Adamo:2017nia}. 
The plane wave `profile' functions $a_\LCperp(x^\LCm)$ are otherwise of arbitrary strength and form. This, combined with the loss of just a single momentum conservation law, leads to amplitudes of significantly greater complexity than those calculated in vacuum. In particular, charged particle propagators and wavefunctions carry a complicated functional dependence (they are `dressed' by the background), and one cannot transform to Fourier/momentum space in any practically useful way.  
To illustrate the typical structures we restrict now to two-point photon amplitudes on the background, which is the focus of this paper. From here on we work in $d=4$, in order to later make contact with spinor-helicity results. 

\medskip

\begin{figure}[t!]
\centering\includegraphics[width=0.8\textwidth]{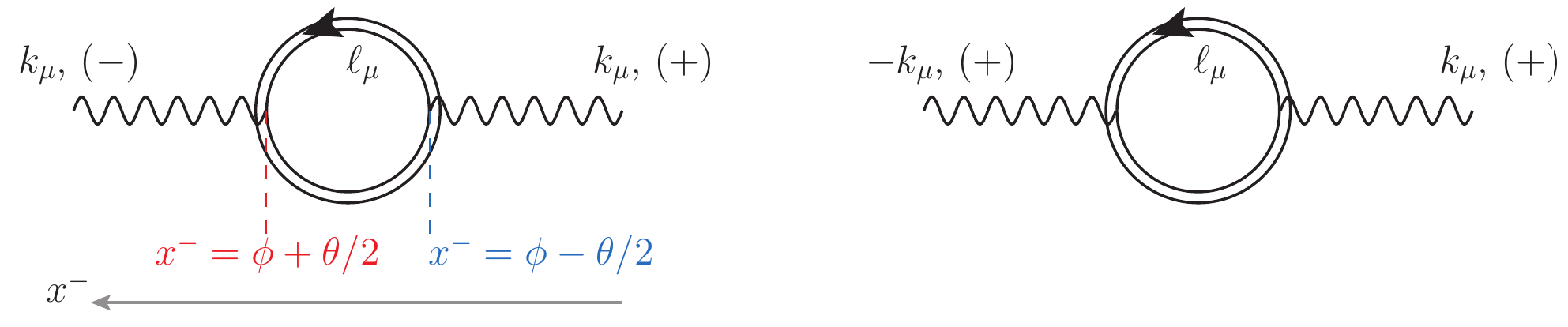}
\caption{\label{FIG:DIAGRAM} 
\textit{Left:} the 1-loop QED amplitude for photon helicity flip on a plane wave background with in/out conventions, i.e.~physical variables.
Double lines indicate the background-dressed fermion propagator. 
The background depends non-trivially on lightfront time $x^\LCm$; our conventions for labelling the vertex positions are shown. 
\textit{Right:} converting to all-incoming conventions, we change the sign of the outgoing photon momentum and swap its helicity state; hence this is the amplitude $\mathcal{M}_{++}$.}
\end{figure}

Consider a two-photon amplitude on a plane wave background. 
These are trivial at tree-level, but \textit{non}-trivial at one-loop and beyond. 
Working in a basis of helicity states for the photons, it is clearly enough to consider the two helicity amplitudes $({+}{+})$ and $({-}{+})$. 
We use all-incoming conventions, so the $({+}{+})$ amplitude describes, physically, \textit{positive to negative helicity flip} of a probe photon traversing the (strong) plane wave background.  Note that \eqref{A-delta-M} then implies that the photon, because it is on-shell, is scattered forward: see Fig.~\ref{FIG:DIAGRAM}.
A compact representation of the corresponding amplitude $\mathcal{M}_{++}$ can be given by first defining the average ${\widehat F}$ of any function $F(x^\LCm)$ of lightfront time by 
\be
{\widehat F } := \frac1\theta \int\limits_{\phi-\theta/2}^{\phi+\theta/2}\!\ud x^\LCm\, F(x^\LCm) \;,
\ee
in which $\phi$ and $\theta$ arise as follows: the one-loop two-point amplitude contains two three-point vertices, meaning two spacetime integrals in the amplitude. $\phi$ and $\theta$ are the average and difference of the two lightfront times of these vertices. See Fig.~\ref{FIG:DIAGRAM}. 
In terms of this average we also define, for any particle of mass $m$, the dressed (`Kibble') mass (cf., \cite{Kibble:1975vz}) which arises in the computation of all amplitudes on plane-wave backgrounds, both in QED and beyond~\cite{Seipt:2017ckc,Adamo:2020qru}:
\be\label{Kibblem}
	M^2(\phi,\theta) := m^2  - \widehat{a^2} +{\widehat a}^{\,2} \equiv m^2 + \text{var}(a)\;,
\ee
which also defines the function `var'. While the Kibble mass has a long history, see~\cite{Harvey:2012ie} and references therein, we do not need its detailed properties -- only that it is quadratic in the background and goes to zero for large argument in any sandwich wave~\cite{Harvey:2012ie}.

We define the dimensionless vector $n_\mu$ by $n\cdot x= x^\LCm = x_\LCp$ for any vector $x^\mu$ (the latter equality follows from (\ref{metric})). Let $\epsilon^{(\LCp)}_\mu$ be the positive helicity polarisation vector of a probe photon, momentum $k_\mu$, and define $\mathcal{A} := \epsilon^{(+)} \cdot {\widehat  a}$. Derivatives of $\cA$ are denoted by subscripts, e.g.~$\mathcal{A}_\theta := \epsilon^{(+)}\cdot \partial_\theta\, {\widehat a}$ and $\mathcal{A}_\phi := \epsilon^{(+)}\cdot \partial_\phi\, {\widehat a}$. Then the $({+}{+})$ amplitude is~\cite{Dinu:2013gaa}

\be\label{simplified}
\mathcal{M}_{++} = -2\im\,\frac{\alpha}{\pi} \int\limits_{-\infty}^\infty\!\ud \phi \int\limits_0^\infty\!\ud\theta\,\theta \, \bigg(\mathcal{A}_\theta^2 -\frac14 \mathcal{A}^2_\phi \bigg)\int\limits_0^1\!\ud s\, \exp\bigg[-\frac{\im}{2n\cdot k} \frac{\theta M^2(\phi,\theta)}{s(1-s)}\bigg]  \;,
\ee
in which $\alpha$ is the fine structure constant, $M^2$ dresses the electron mass as in (\ref{Kibblem}), and the variable $s$ arises as the `lightfront momentum fraction' of the loop electron momentum~$l_\mu$, specifically $s= n\cdot l / n \cdot k$. (The transverse loop degrees of freedom have been integrated out and the  amplitude is UV finite.) Note that 2-point scattering is forward on-shell because 3 momentum components are conserved -- see \eqref{A-delta-M} -- so no particle labels are needed on polarizations.
The integral over $s$ can be performed to yield Bessel functions~\cite{Dinu:2013gaa}, while those over the lightfront times $\theta$ and $\phi$ can only be performed analytically for a few special cases. 
The total probability of helicity flip is $\mathbb{P} = |\mathcal{M}_{++}|^2 / (n\cdot k)^2$~\cite{Dinu:2013gaa,Adamo:2019zmk}.

The fact that ${\mathcal M}_{++}$ is \textit{non}-vanishing yields macroscopic phenomena such as vacuum birefringence~\cite{Toll:1952rq,Heinzl:2006xc} (reviewed in~\cite{King:2015tba}), where a beam of photons traversing a strong background develops an ellipticity in its polarisation due to helicity flip of the constituent photons. 
This ellipticity is directly related to the helicity flip amplitude~\cite{Dinu:2013gaa}.

We turn now to the one-loop $({-}{+})$ amplitude. 
Physically, this describes forward scattering (no helicity flip) of a probe photon. 
In contrast to the corresponding one-loop amplitude in vacuum, this expression contains non-trivial physics beyond renormalisation. 
The real part of the amplitude gives the `vacuum refractive indices' which form the low-energy description of photon-photon scattering. 
At higher energies these indices become complex, exhibiting both dispersive and absorptive behaviour~\cite{Shore:2007um}. 
In other words, the amplitude develops an imaginary part equal to the total (tree-level) probability of electron-positron pair production from a probe photon entering the plane wave\footnote{In general, the helicity flip amplitude $\mathcal{M}_{++}$ also has an imaginary part. However, the interpretation of this is less clear, since (in the in-out picture) the probe photons have opposite helicity.}, see e.g.~\cite{Nikishov:1964zza,Heinzl:2010vg,Meuren:2013oya}.

Defining the new quantity $\bar{\mathcal{A}} := {\epsilon}^{(-)}\cdot {\widehat a} $, the amplitude ${\mathcal M}_{-+}$ is
%
%
%
%
\be\label{simplify-me}
\begin{split}
    \mathcal{M}_{-+} =  -2\im\,\frac{\alpha}{2\pi}
    \int\limits_{-\infty}^\infty\!\ud \phi
    \int\limits_0^\infty\!\ud\theta\,\theta \!\int\limits_0^1\!\ud s\, 
    &
    e^{-\frac{\im}{2n\cdot k} \frac{\theta M^2}{s(1-s)}} 
    \bigg[
        {\epsilon}^{(-)}\cdot{\epsilon}^{(+)} 
        \bigg(
            \frac{M^2-m^2}{\theta^2 M^2}
            \partial_\theta(\theta M^2) 
            + 
            \frac{{\widehat a}_\phi \cdot {\widehat a}_\phi}{4s(1-s)}
        \bigg)
        \\
        &
        -
        \frac12 
        {\bar{\mathcal{A}}}_\phi {{\mathcal{A}}}_\phi 
        + 2 {\bar{\mathcal{A}}}_\theta {{\mathcal{A}}}_\theta 
        - 
        \bigg(
            1 - \frac{1}{2s(1-s)}
        \bigg)
        {\bar{\mathcal{A}}}_{[\theta}{{\mathcal{A}}}_{\phi]}
    \bigg] \;,
\end{split}
\ee
where the square brackets in $\bar{\mathcal{A}}_{[\theta}\mathcal{A}_{\phi]}$ denote anti-symmetrization (without a factor of $1/2$). 
This has a more complicated form than ${\mathcal M}_{++}$ largely due to the first term in large square brackets, which results from renormalisation of the usual one-loop UV divergence, see~\cite{Dinu:2013gaa} for details. 
Analogous amplitudes can be found in Yang Mills and QCD on plane wave backgrounds~\cite{Adamo:2019zmk}, as will be discussed below.


\subsection{Access to higher-point amplitudes}\label{subsec:access}

Abelian plane waves are characterised by the same functional degrees of freedom, in both form and number, as the on-shell degrees of freedom of a photon.
(Analogous statements hold for Yang-Mills plane waves and gluons, see below).  
As such, plane wave backgrounds can be viewed as coherent superpositions of gauge bosons; this is made clear as follows. Let $S[eA]$ denote the standard QED $S$-matrix, in which we highlight, for what follows, the dependence on the quantised gauge field~$A_\mu$. 
Consider a scattering amplitude in which the initial and final states have, as well as some number of scattered particles, the same coherent state of photons, $\ket{z}$. 
Using the representation of the coherent states in terms of a displacement operator, one can show that~\cite{KibbleShift,Frantz}
\be\label{relat1}
	\bra{\text{out},z} S[eA ]\ket{\text{in},z} = 		\bra{\text{out}} S[eA + a]\ket{\text{in}} \;,
\ee
such that scattering between the coherent states is equivalent to scattering in a background field $a = eA_\text{bg}$ obeying Maxwell's equations in vacuum, $\partial_\mu F_\text{bg}^{\mu\nu}=0$. In terms of Fourier modes the background is
\begin{align}\label{eqn:Fourier}
    a_{\mu}(x)
    =
    \int \!\frac{\ud^3 {\bf k}}{(2\pi)^32k_0}\,
    \Big(
        z_{\mu}(k)
        e^{-\im k \cdot x}
        +
        \bar{z}_{\mu}(k)
        e^{+\im k \cdot x}
    \Big) 
    \,.
\end{align}
%
The equality \eqref{relat1} holds up to contact terms; we assume that the particles implicit in $\ket{\text{in}}$ and $\bra{\text{out}}$ are scattered into regions of momentum space distinct from that on which the background has support -- for details, subtleties and extensions see~\cite{KibbleShift,Frantz,Gavrilov:1990qa,Ilderton:2017xbj}.  Now, since a coherent state is itself a superposition of states of definite photon number,
\be
	\ket{z} = \sum\limits_{N=0} \frac{z^N}{\sqrt{N!}} \ket{N}
\ee
it follows that if $z$ is characterised by some amplitude $a_0$, then the coefficient of  $a_0^N$ in the amplitude on background, $\bra{\text{out}} S[eA + a]\ket{\text{in}}$, is proportional to the sum of all vacuum amplitudes including $N$ photons in all possible incoming/outgoing configurations, weighted with numerical factors and convoluted with the field profile~\cite{Seipt:2010ya,Ilderton:2019bop,Adamo:2020qru}. As a sandwich plane wave of the form \eqref{eA} has compact support, this is the same as saying that the vacuum amplitudes contained in $\bra{\text{out}} S[eA + a]\ket{\text{in}}$ describe scattering between photon \textit{wavepackets}. Thus, to recover the vacuum limit of `standard' amplitudes via a perturbative expansion in the background, one must strip off the wavepacket such that the background photons are treated as single Fourier modes. We will show below how this can be implemented through appropriate choices of field profile, allowing us to extract any particular $(N-r)$-point to $r$-point vacuum amplitude.

%% file: QED.tex

It was established long ago that for photon scattering in massless QED, the helicity configurations $(\pm +\cdots +)$ have vanishing one-loop amplitudes unless the total number of external photons is exactly four~\cite{Mahlon:1993fe}. 
In the four-point case, the resulting amplitudes are rational functions of the momenta, with extremely compact expressions in the spinor helicity formalism. 
In a sense, this is not surprising since the all-plus configuration arises from the purely self-dual sector of the theory, which is classically integrable, and the one-minus-rest-plus configuration is a first perturbation away from this integrable subsector (cf., \cite{Bardeen:1995gk,Cangemi:1996rx}). 
While these four-point amplitudes are not rational in massive QED, their rational part is the same as the massless case (cf., \cite{Bernicot:2008th}).

In this section we consider the perturbative expansion of the 2-point, 1-loop amplitudes of strong field QED in a \emph{self-dual} plane wave background. 
Self-duality ensures that the perturbative expansion generates only positive helicity photons from the background. 

\subsection{Self-dual plane wave backgrounds}

A generic electromagnetic plane wave can be decomposed as
\be\label{gpw}
a_{\mu}=\frac{f(x^\LCm)}{\sqrt{2}}\,(0,0,1,\im) +  \frac{{\bar f}(x^\LCm)}{\sqrt{2}}\,(0,0,1,-\im) \;,
\ee
where $f(x^-)$ and $\bar{f}(x^-)$ are related by complex conjugation for Lorentzian-real gauge fields. 
In particular, $f$ can be thought of as encoding the positive helicity portion of the plane wave, while $\bar{f}$ encodes the negative helicity portion. 
For \emph{complex} fields, $f$ and $\bar{f}$ become independent functions of lightfront time, and we can set $\bar{f}=0$ while keeping $f\neq 0$. 
The resulting chiral plane wave is a \emph{self-dual plane wave} (SDPW), since its field strength is a self-dual 2-form (with respect to Hodge duality on Minkowski space), and, from the above,  can be viewed as a coherent superposition of positive helicity photons.

For such a SDPW, the potential $a_{\mu}$ obeys $a^2=0$ and can be represented in the 2-spinor formalism as: 
\be\label{EMgf}
 a_\mu(x^\LCm) \leftrightarrow \,\, \sqrt{2} f(x^-) \, \frac{\ket{o}\sbra{\iota}}{\braket{{o}\,\iota}}\;,
\ee
in which $|\iota\ra$, $|o\ra$ is a spinor dyad obeying $\la \iota\,o\ra=1$, with $\ket{\iota}\sbra{\iota}$ being the spinor representation of the background direction $n_\mu$. (Our spinor helicity conventions are that $\langle 12\rangle [21] = 2 p_1 \cdot p_2$.)  
The background will generate only positive helicity photons, giving more control over the helicity configurations that are generated by the perturbative expansion at one-loop. 
Two-point amplitudes on SDPWs have not previously been considered in the literature, but they are easily extracted from corresponding results in real plane waves by analytically continuing the background and imposing~\eqref{EMgf}.
    
\subsection{Collinear limits of all-plus amplitudes $({+}{+}{+}{+})$}
Consider the 2-point amplitude $\cM_{++}$ of~\eqref{simplified}. 
Restricted to a SDPW background, the perturbative expansion of $\cM_{++}$ will generate 1-loop vacuum amplitudes with \textit{only} positive helicity external photons: two of these will be the probe photons from the initial 2-point amplitude, with the rest being extracted from the SDPW background. 

In massless QED, there is only one such all-plus helicity amplitude at 1-loop which is non-vanishing in vacuum; this is the 4-point amplitude~\cite{Mahlon:1993fe,Gastmans:1990xh}
\be\label{4pap}
A^{(1)}_{4}(\gamma_1^{+},\gamma_2^{+},\gamma_3^{+},\gamma_4^{+})=(2\pi)^4\,\delta^{4}\!\left(\sum_{i=1}^{4}k_{i}\right)\,\frac{\im\,e^4}{2\,\pi^2}\, \frac{{\sbraket{12}}\,{\sbraket{34}}}{\braket{12}\,\braket{34}}\,.
\ee
All other $({+}{+}\cdots {+})$-helicity photon amplitudes vanish at 1-loop. 
The implication for the perturbative expansion of $\cM_{++}$ on SDPW background is clear: it must truncate at four external photons. 

\medskip
 
Now, for a free photon with momentum $k_\mu$, it is convenient to take the reference spinor in its polarisation vector to be aligned with the background direction $n_\mu$, so that positive and negative helicity states are represented by
    \be
    \epsilon^{(\LCp)}_{\mu} \,\, \leftrightarrow \,\, \sqrt{2}\,\frac{|\iota\ra\,[k|}{\la\iota\,k\ra}\;,
    \qquad \qquad
      \epsilon^{(\LCm)}_{\mu} \,\, \leftrightarrow \,\, \sqrt{2}\,\frac{|k\ra\,[\iota|}{[\iota\,k]} \;.
     \ee
This corresponds to imposing a hybrid Lorenz-lightfront gauge, so that $k\cdot\epsilon^{(\pm)}=0=n\cdot\epsilon^{(\pm)}$. 

These explicit spinor representations can then be fed directly into the helicity flip amplitude $\cM_{++}$ of \eqref{simplified}. 
First, observe that for a SDPW background of the form \eqref{EMgf}, the quantity $\mathrm{var}(a)$ which enters the Kibble mass vanishes since both $\widehat{a}^{2}$ and $\widehat{a^{2}}$ are zero, leaving only the bare electron mass in the exponential part of the amplitude integrand. 
In the pre-exponential portion of the amplitude, one finds
    \begin{equation}
        \mathcal{A} = \epsilon^{(\LCp)}\cdot \widehat{a} = - \widehat{f}\, \frac{\sbraket{\iota\,k}}{\braket{\iota\,k}} \;,
    \end{equation}
using the properties of the spinor dyad.

With this, the amplitude can be factored into a `vacuum' part which contains the spinors and is un-dressed by the background, and a background-dependent part:    
    \be\label{simplified3}
\mathcal{M}_{++} \to -2\im\,\frac{\alpha}{\pi}\, \int\limits_{-\infty}^\infty\!\ud \phi \int\limits_0^\infty\!\ud\theta\,\theta \int\limits_0^1\!\ud s\, \e^{-\frac{\im}{2n\cdot k} \frac{\theta\, m^2}{s(1-s)}}  \, \Big({{\widehat f}_\theta}^{\, 2} -\frac14 {{\widehat f}_\phi}^{\, 2} \Big)\frac{{\sbraket{\iota\,k}}^2}{\braket{\iota\,k}^2} \;.
\ee
Recalling that vacuum amplitudes are extracted by repeated functional differentiation with respect to the background, we see immediately that the perturbative expansion of \eqref{simplified3} cannot generate vacuum amplitudes with more than four external legs. 
This is because \eqref{simplified3} has only pre-exponential dependence on the background, which is quadratic in the profile function $f$. 
In other words, self-duality of the background kills all non-polynomial dependence on the background, ensuring truncation of the perturbative expansion, as required.

\begin{figure}[t!]
\centering \raisebox{40pt}{$\mathcal{M}_{++} \longrightarrow A^{(1)}_4(++++) $}\qquad\includegraphics[width=0.45\textwidth]{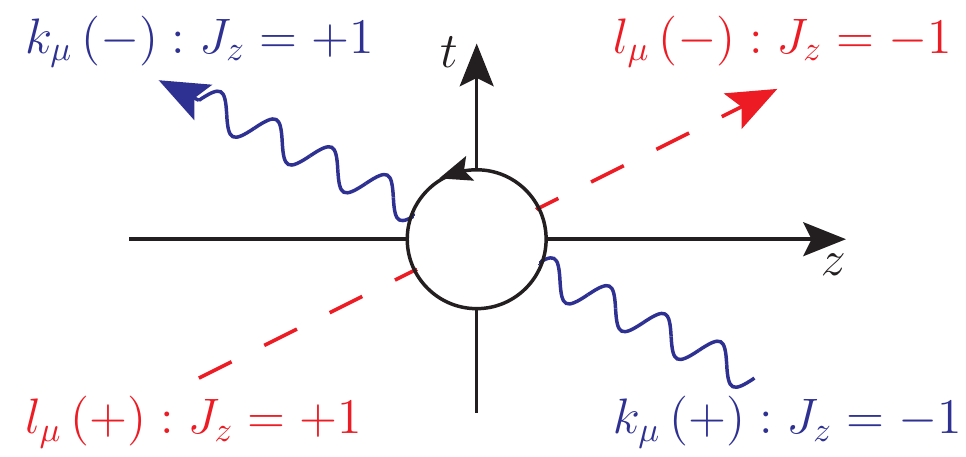}
\caption{\label{FIG:ALLPLUS} The perturbative expansion of $\mathcal{M}_{++}$ yields the four-point ({+}{+}{+}{+}) amplitude, here re-expressed in physical (in/out) variables; the blue/wavy line indicates the probe photon, the red/dashed line indicates the background photon. Momentum, helicity and angular momentum~$J_z$  are given for each, the latter assuming a head-on collision. Space-time directions are included to allow easier identification of the physical $J_z$ assignments.}
\end{figure}

Explicit comparison with the vacuum amplitude \eqref{4pap} is now possible by setting the electron mass to zero and evaluating all integrals. 
To do this, we decompose the background profile into linear incoming and outgoing modes of frequency $\omega$ (recall $x^\LCm = n\cdot x$),
\be\label{bdecomp}
f(x^\LCm) = e\, c_\text{in}\, \e^{-\im\omega\, n\cdot x} + e\, c_\text{out}\, \e^{\im\omega\, n\cdot x}\,,
\ee
and extract the coefficient of $c_\text{in}\,c_\text{out}$, so that one of the background photons is incoming and the other outgoing; see Fig.~\ref{FIG:ALLPLUS}. In physical variables, \textit{both} the probe and background photons undergo helicity flip: note that while the tensor structure of the SDPW background contains only positive helicity polarisations, the physical meaning of these polarisations depends on whether the corresponding photon is in the final state or initial state.

After computing the relevant coefficient of
$c_\text{in}c_\text{out}$ one finds that \eqref{simplified3} becomes:
\be\label{simplified4}  
\mathcal{M}_{++} = -2\im\,\frac{\alpha}{\pi}\,\int\limits_{-\infty}^\infty\!\ud \phi \int\limits_0^1\!\ud s \int\limits_0^\infty\!\ud t\, 
\e^{\frac{-\im\,t\, m^2}{2\omega\, n\cdot k\, s(1-s)}}
\left(\frac{2 \left(t^2-2\right) \cos t-4 t\sin t+4}{t^3}\right)
\frac{{\sbraket{\iota\,k}}^2}{\braket{\iota\,k}^2} \;,
\ee
where we have changed variables from $\theta$ to dimensionless $t$ defined by $\theta \to t = \theta/\omega$. 

We comment further on the massive case in Appendix \ref{app:MASS}, but for now set $m=0$; all three remaining integrals factorize, since the ratio of spinor brackets is independent of the loop momentum and position space variables. 
The $s$ integral gives a factor of $1$, while the $t$ integral gives a factor of $-1$. 
The integral over $\phi$ gives the volume of lightfront time; reintroducing the three-momentum delta function stripped off in \eqref{A-delta-M}, and writing $k'_\mu$ for the final state photon momentum,  the volume factor may be written as $2\pi \delta(0) = 2\pi \delta(k'_\LCm + k_\LCm)$,  recovering full four-momentum conservation. 
Thus, combining \eqref{simplified4} with \eqref{A-delta-M} leaves
\be\label{ppfinal}
		(2\pi)^4\, \delta^4(k'+k+\omega n-\omega n) \, \frac{\im\,e^4}{2\,\pi^2} \,\frac{{\sbraket{\iota\,k}}^2}{\braket{\iota\,k}^2} \;,
\ee
where the incoming and outgoing momenta $\pm \omega n_{\mu}$ of the background photons has been trivially included in the overall momentum conserving delta function. 

To compare with the all-positive helicity vacuum amplitude \eqref{4pap}, we must take a doubly-collinear limit of that amplitude since, in our SDPW calculation, the two photons from the background are collinear and (by momentum conservation) the `probe' photons are collinear. 
Sure enough, in this limit
\begin{equation}
\lim_{\substack{1,3\to \omega n \\ 2,4\to k}}\,\,\frac{\im\,e^4}{2\,\pi^2}\, \frac{{\sbraket{12}}\,{\sbraket{34}}}{\braket{12}\,\braket{34}} = \frac{\im\,e^4}{2\,\pi^2} \,\frac{{\sbraket{\iota\,k}}^2}{\braket{\iota\,k}^2} \;,
\end{equation}
in precise agreement with \eqref{ppfinal}. 
Thus we recover the correct double collinear limits of the rational all-plus amplitude in massless QED. 

The calculation in massive QED is slightly more subtle, as the vacuum amplitude is no longer a rational function. However, it can be shown that \eqref{simplified4} with $m\neq0$ correctly reproduces the double collinear limit of the massive vacuum amplitude -- this calculation is detailed in Appendix~\ref{app:MASS}. In Appendix~\ref{app:six-point} we provide a further check on our results, by expanding $\cM_{++}$ of \eqref{simplified} to higher orders in a \textit{non-chiral} plane wave background, and matching to a six-point vacuum amplitude.

A further check on the multicollinear limits of vacuum amplitudes (both here and below) is provided by angular momentum, $J_z$, conservation in the case that the initial state photons collide head-on with the background (cf., \cite{Gaunt:2011xd,Dixon:2016epj} for applications of this argument in the context of double parton scattering). The $J_z$ assignments of initial and final state photons are most easily seen with in/out conventions, and these are shown in Fig.~\ref{FIG:ALLPLUS} for the four-point vacuum amplitude derived from $\mathcal{M}_{++}$: angular momentum is conserved in this case, so a non-vanishing result is allowed.


\subsection{Collinear limits of one-minus-rest-plus amplitudes $({-}{+}{+}{+})$}
\begin{figure}[t!]
\centering \raisebox{40pt}{$\mathcal{M}_{-+} \longrightarrow A^{(1)}_4(-+++) $}\qquad\includegraphics[width=0.55\textwidth]{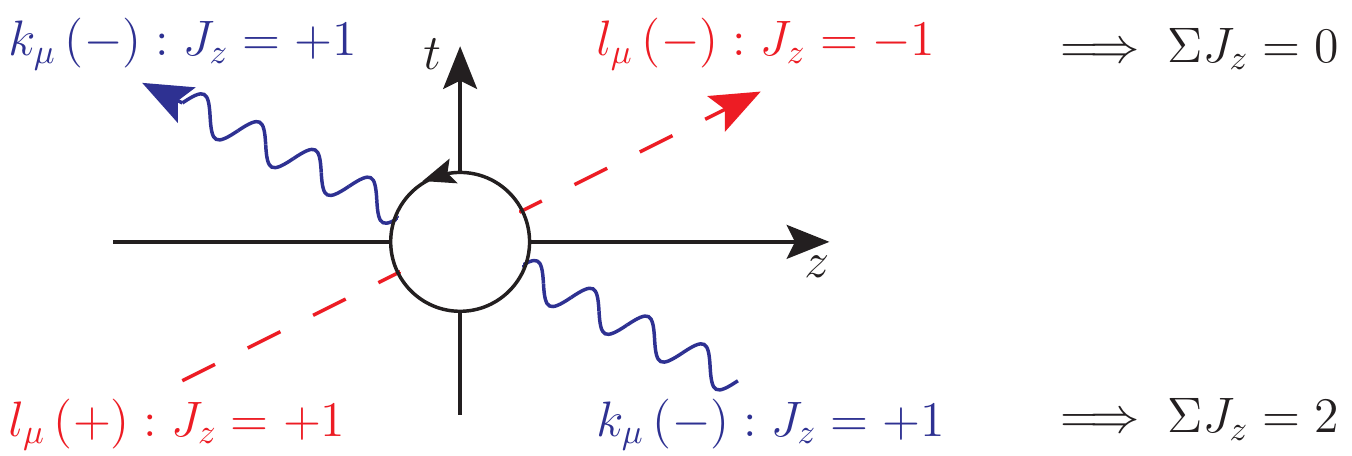}
\caption{\label{FIG:RESTMINUS}
The perturbative expansion of $\mathcal{M}_{-+}$ yields the four-point ({--}{+}{+}{+}) amplitude, here again re-expressed in terms of physical (in/out) variables. Notation is as in Fig.~\ref{FIG:ALLPLUS}. Note that angular momentum is not conserved in the limit that the initial state photons collide head on, and indeed the vacuum amplitude is found to be zero
}
\end{figure}

Starting from $\cM_{-+}$ in a SDPW background, perturbative expansion will generate one-loop photon amplitudes in vacuum with a single negative helicity photon and the remainder positive helicity; see Fig.~\ref{FIG:RESTMINUS}.
Much like the all-positive helicity case, the only non-vanishing one of these vacuum amplitudes in massless QED is at four points~\cite{Gastmans:1990xh}:
\be\label{4p1n}
A^{(1)}_{4}(\gamma_1^{-},\gamma_2^{+},\gamma_3^{+},\gamma_4^{+})=(2\pi)^4\,\delta^{4}\!\left(\sum_{i=1}^{4}k_{i}\right)\,\frac{\im\,e^4}{2\,\pi^2}\, \frac{\braket{12}\,{\sbraket{24}}\,\braket{41}}{\braket{23}\,\braket{34}\,\braket{24}}\,,
\ee
so we should find that the perturbative expansion of $\cM_{-+}$ truncates at four external photons.

Indeed, this property is trivially satisfied upon restricting \eqref{simplify-me} to a SDPW background, even before performing the perturbative expansion: in a SDPW, $M^2=m^2$, $a^2=0$ (and hence $\widehat{a}^2_{\phi}=0$), and $\bar{\mathcal{A}}=0$ because $\bar{f}=0$.
In other words, the helicity non-flip amplitude actually \emph{vanishes} identically in an SDPW background,
\be\label{nf1}
\cM_{-+}\to0\,.
\ee
This is of course consistent with the vanishing one-minus-rest-plus amplitudes for five or more photons. 
For the four-point amplitude, consistency requires that the double collinear limit of the vacuum amplitude \eqref{4p1n} should vanish. 
However, taking photons 1 and 3 to be collinear with $k$ and 2 and 4 collinear with $n$ in \eqref{4p1n} appears ill-defined, producing a 0/0 result. 
This merely signals that further care is required when taking the double collinear limit in this helicity configuration.   
The limit can be regularized by taking~\cite{Stieberger:2015kia,Nandan:2016ohb,Bhattacharya:2018vph}
\be\label{dcreg}
|1\ra\rightarrow |k\ra\,, \quad |3\ra\rightarrow |k\ra+\varepsilon\,\frac{|\iota\ra}{\la k\,\iota\ra}\,, \quad |2\ra\rightarrow |\iota\ra\,, \quad |4\ra \rightarrow |\iota\ra+\varepsilon\,|o\ra\,,
\ee
and similarly for the square bracket spinors, so that in the (almost) collinear limit
\be\label{dcreg2}
\la 24\ra\,, \;[24]\rightarrow \varepsilon\,, \qquad \la13\ra\,, \; [13]\rightarrow \varepsilon\,, 
\ee
and in the end the regulator $\varepsilon\rightarrow 0$.

The double collinear limit can be further clarified by exploiting a relation equivalent to the photon-decoupling identity in non-abelian gauge theory (cf., \cite{Mangano:1990by,Bern:1990ux,Bern:1991aq}):
\begin{multline}\label{dcreg3}
\frac{\im\,e^4}{2\,\pi^2}\, \frac{\braket{12}\,{\sbraket{24}}\,\braket{41}}{\braket{23}\,\braket{34}\,\braket{24}}= \\
\frac{\im\,e^4}{6\,\pi^2}\left(-\frac{{\sbraket{24}}^3\,\braket{24}}{{\sbraket{12}}\,\braket{23}\,\braket{34}\,{\sbraket{41}}} +\frac{{\sbraket{34}}^3\,\braket{34}}{{\sbraket{13}}\,\braket{23}\,\braket{24}\,{\sbraket{41}}}+\frac{{\sbraket{23}}^3\,\braket{23}}{{\sbraket{12}}\,\braket{24}\,\braket{34}\,{\sbraket{31}}}\right)\,.
\end{multline}
Written in this fashion, the double collinear limit as defined by \eqref{dcreg} becomes
\be\label{dcreg4}
\frac{\im\,e^4}{6\,\pi^2}\,\lim_{\varepsilon\to0}\left(\frac{-\varepsilon^4}{{\sbraket{\iota\,k}}\,\braket{\iota\,k}}+\frac{{\sbraket{\iota\,k}}^2}{\varepsilon^2}-\frac{{\sbraket{\iota\,k}}^2}{\varepsilon^2}\right)=0\,,
\ee
as the last two terms cancel prior to the $\varepsilon\to0$ limit. 
Thus, the vanishing of $\cM_{-+}$ in a SDPW background is consistent with the double collinear limit of the four-photon amplitude \eqref{4p1n}.

Note that $\cM_{-+}$ vanishes on a SDPW background for any value of the electron mass. 
For massive QED, the $({-}{+}{+}{+})$ vacuum amplitude is given by a sum over terms, each of which is proportional to (a permutation of) \eqref{4p1n} multiplied by a scalar integral (cf., \cite{Bern:1995db,Bernicot:2008th}). 
Thus, the doubly collinear limit of the massive amplitude also vanishes, as required.

\subsection{Split helicity $({-}{-}{+}{+})$}
Unlike the $({+}{+}{+}{+})$ and $({-}{+}{+}{+})$ helicity configurations, the one-loop split helicity amplitude $({-}{-}{+}{+})$ is not a rational function of kinematic data in massless QED. 
This is the only remaining helicity configuration at four-points, so it is natural to ask if its double collinear limits can be recovered from two point amplitudes on a SDPW background.  
As the background can only contribute positive helicity photons, the split-helicity configuration must arise from the perturbative expansion of $\cM_{--}$ on the background.
	
Recall that $\cM_{--}$ is obtained from $\cM_{++}$ by simply replacing the probe polarization $\epsilon^{(+)}$ with $\epsilon^{(-)}$. 
At the level of \eqref{simplified}, this entails replacing $\cA$ with $\bar{\cA}$ in the pre-exponential of the integrand. 
However, on a SDPW background $\bar{\cA}=0$ and thus $\cM_{--}=0$.
This must be reflected in the double collinear limit of the appropriate split helicity vacuum amplitude\footnote{That $\cM_{++}\neq 0$ and $\cM_{--}=0$ are not simply parity conjugates is, of course, explained by the fact that the SDPW background is chiral, and thus breaks parity.}. 
	
\begin{figure}[t!]
\centering \raisebox{40pt}{$\mathcal{M}_{--} \longrightarrow A^{(1)}_4(--++) $}\qquad\includegraphics[width=0.55\textwidth]{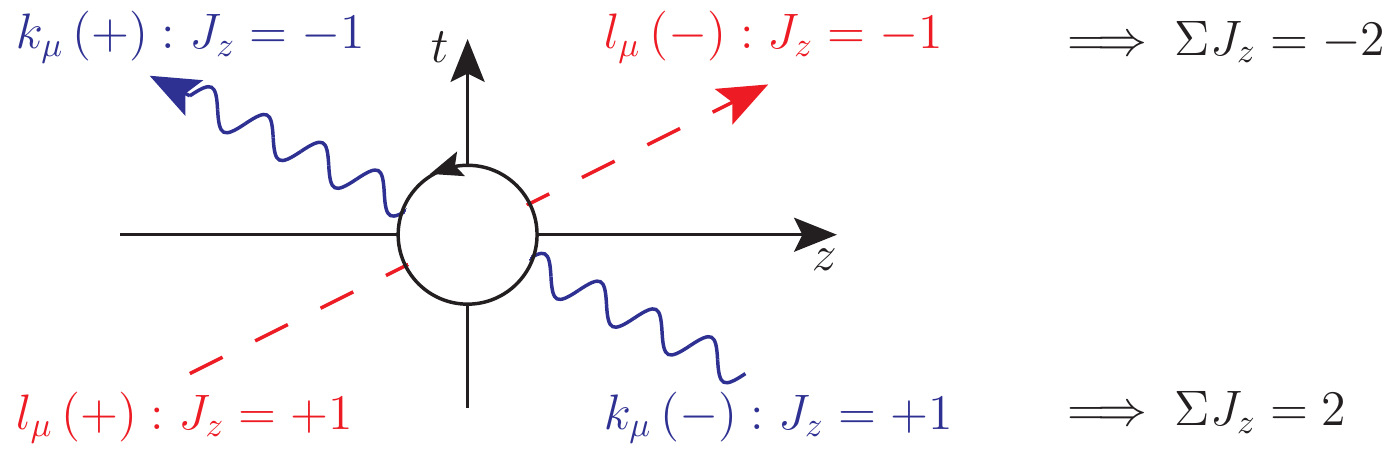}
\caption{\label{FIG:SPLIT} Expanding $\mathcal{M}_{--}$ yields the split helicity amplitude $A_4^{(1)}(--++)$. The corresponding physical situation is negative-to-positive helicity flip of a probe photon on the SDPW background, expansion of which generates a four-point amplitude in which the initial state photons have opposite helicity, and both undergo helicity flip. That $\mathcal{M}_{--}$ vanishes means that there is no negative-to-positive helicity flip on a SDPW background. Again, this is consistent with (the violation of) angular momentum conservation, as shown.
}
\end{figure}

To identify this amplitude, we must briefly return to in/out variables. Recall that for four-point scattering in vacuum, the only interesting cases are $2\to2$ scattering (cf., Appendix B of~\cite{Gastmans:1990xh}). 
If $h_i$ denotes the helicity of particle $i$ and particles 1,2 are incoming and 3,4 are outgoing, then configurations are denoted by $A_{4}(h_1\, h_2;h_3\,h_4)$. 
The possible split helicity configurations are thus $A_{4}({+}{+};{-}{-})$, $A_{4}({-}{-};{+}{+})$ or $A_{4}({+}{-};{+}{-})$. 
In $\mathcal{M}_{--}$, the incoming probe photon has negative helicity, and the \emph{outgoing} probe has positive helicity. 
Perturbatively expanding the positive helicity background yields an additional positive (negative) helicity photon in the in-state (out-state); hence the amplitude of interest must be $A_{4}({+}{-};{+}{-})$, in which momenta $k_1$ and $k_4$ come from the background\footnote{The remaining cases describe highly degenerate scenarios in which the background photons appear only in the in-state or out-state (and momentum conservation dictates that \emph{all} the photons have collinear momenta), corresponding to vacuum emission or absorption of photons, which is zero in a plane wave.}; see Fig.~\ref{FIG:SPLIT}.
 %

Recalling that we use in/out conventions for a moment, the one-loop vacuum amplitude for $A_{4}(+-;+-)$ in massless QED
is~\cite{Gastmans:1990xh}
\begin{align}\label{eqn:MHVPhoton}
    A^{(1)}_{4}(\gamma_1^{+},\gamma_2^{-};\,\gamma_3^{+},\gamma_4^{-})
    =
    &
    \frac{e^{4}}{2\, \pi^{2}}
    \left[
        \frac{s^{2} + t^{2}}{2 u^{2}}
        \ln^{2}\!\left(
            \frac{-t}{s}
        \right) 
        -
        \frac{s - t}{u}
        \ln\left(
            \frac{-t}{s}
        \right) 
        +
        1 \right.
        \nonumber\\
        &
        \qquad
        \left.+
        \im \pi
        \left(
            \frac{s^{2} + t^{2}}{u^{2}}
            \ln\left(
                \frac{-t}{s}
            \right) 
            -
            \frac{s - t}{u}
        \right)
    \right]\,,
\end{align}
where we have omitted the overall momentum-conserving delta function and $s = 2 k_{1} \cdot k_{2}$, $t = - 2 k_{1} \cdot k_{3}$ and $u = - 2 k_{1} \cdot k_{4}$ are the usual Mandelstam invariants, with $s + u + t = 0$. 
Given the discussion above, we take the limit in which $k_1$ and $k_4$, corresponding to background photons, become collinear (i.e., $k_1,k_4\to n$), while $k_1\cdot k_2\to n\cdot k$ remains fixed. 
Setting $\varepsilon = 2k_1\cdot k_4$, this corresponds to the regime where $s$ is fixed and non-zero, while $u= - \varepsilon$ and $t= -s + \varepsilon$ in the $\varepsilon \to 0$ limit. 
A straightforward -- if tedious -- calculation gives
\be
	A^{(1)}_{4}(\gamma_1^{+},\gamma_2^{-};\,\gamma_3^{+},\gamma_4^{-})
	\rightarrow
        -\im\,\frac{e^{4}}{3 \pi}\, \frac{\varepsilon}{s}+O(\varepsilon^2)\,,
\ee
which vanishes as $\varepsilon\to 0$, as required for consistency with the vanishing of $\cM_{--}$ on the SDPW background.


%% file: YangMills.tex
%

For non-abelian Yang-Mills theory, there are non-vanishing one-loop gluon scattering amplitudes in vacuum for the helicity configurations $({\pm}{+}\cdots{+})$ to all multiplicity. 
These are constrained to be rational functions since their cuts produce tree-amplitudes which vanish in the relevant helicity configurations, and are given by remarkably compact all-multiplicity formulae~\cite{Bern:1993qk,Mahlon:1993si,Bern:1994ju,Bern:2005ji}. 
Recalling the discussion of Section~\ref{sec:QED}, this simplicity can again be understood as arising from the integrability of the self-dual sector of Yang-Mills theory (cf., \cite{Bardeen:1995gk,Cangemi:1996rx,Bern:1996ja}). 

As in QED, we expect to be able to access highly collinear configurations of these rational one-loop amplitudes by perturbatively expanding helicity-flip and helicity-non-flip two-point amplitudes on a self-dual plane wave background.  
One could worry that the collinear singularities of Yang-Mills theory could produce divergent results, but the fact that all background gluons are Cartan-valued means that the resulting vacuum amplitudes are in fact equivalent to mixed gluon-photon amplitudes, whose structure is tightly constrained~\cite{Bern:1993qk}. 


\subsection{Self-dual plane waves \& dressed spinor-helicity variables}

In Yang-Mills theory, a plane wave solution has the same Heisenberg symmetry algebra as in the abelian setting of QED. 
These symmetries constrain the gauge field to be valued in a Cartan subalgebra of the gauge group~\cite{Trautman:1980bj,Adamo:2017nia}:
\be\label{ympw}
a_{\mu}=\frac{f(x^\LCm)}{\sqrt{2}}\,(0,0,1,\im) +  \frac{{\bar f}(x^\LCm)}{\sqrt{2}}\,(0,0,1,-\im)\,,
\ee
where $f,\,\bar{f}$ are Cartan-valued functions, related by complex conjugation for Lorentzian-real fields. 
Thus, a self-dual plane wave in Yang-Mills theory is defined by setting $\bar{f}=0$ as in \eqref{EMgf}; the only difference from QED is that $f(x^-)$ is now Cartan-valued.

By virtue of being Cartan-valued, Yang-Mills plane waves are essentially abelian in nature. 
Nevertheless, there are important differences between QED and Yang-Mills theory in plane wave backgrounds. For instance, gluons propagating -- carrying generic non-abelian colour -- are dressed by the background, in contrast to photons~\cite{Adamo:2017nia}. 
One-particle gluon states can be given explicitly, and their kinematic data (i.e., momenta and polarisations) are also dressed. 
Fortunately, a dressed version of the spinor helicity formalism enables a compact description of the dressed kinematics~\cite{Adamo:2019zmk,Adamo:2020syc}.

For a gluon with initial on-shell momentum $k=\ket{k}\sbra{k}$, the dressed momentum in a SDPW background is
\be\label{glumom}
K=\ket{k}\,[\!\sbra{K}\,, \qquad
\sket{K}\!]:=\sket{k}-\frac{ef(x^-)}{\sqrt{k_\LCp}}\,\sket{\iota}\,,
\ee
where double-bracket notation denotes a spinor dressed by the background. 
Here, $e$ is the `charge' of the gluon perturbation with respect to the SDPW background, defined as follows. Let $a_\mu = a_\mu^\sfi \mathsf{t}^{\sfi}$ for $\mathsf{t}^\sfi$ the generators of the Cartan. Then if $\mathsf{T}^\mathsf{a}$ is the colour vector of the gluon, valued in a root eigenspace of the Cartan, its charge is defined by $[\mathsf{t}^\sfi, \mathsf{T}^\mathsf{a}] = e^\mathsf{i} \mathsf{T}^\mathsf{a}$. It follows that $[a_\mu,\mathsf{T}^\mathsf{a}]=e^\mathsf{i} a_\mu^\sfi \mathsf{T}^a$. 
The contraction of explicit colour indices between $e$ and the background is always suppressed, so we write e.g.~$e^\mathsf{i} a^\sfi_\mu = e a_\mu$ (and similarly for $f$ in \eqref{ympw}). Note that we sometimes abuse notation by separating out the charge, writing e.g.~$e^2 a^2$ instead of $(ea)^2$.

Chirality of the SDPW background ensures that only one of the initial momentum spinors is dressed, namely the positive helicity one: $\sket{k}\to\sket{K}\!]$. 
As a result, the polarization of positive helicity gluons is dressed in a SDPW while negative helicity gluons remain un-dressed:
 \be\label{GLUON-POL-DRESSED}
    \cE^{(\LCp)}(x^\LCm) = \sqrt{2}\,\frac{\ket{\iota}\,[\!\sbra{K}}{\braket{\iota\,k}} \;,  
    \qquad \qquad
      \cE^{(\LCm)}(x^\LCm) \equiv \epsilon^{(-)} = \sqrt{2}\,\frac{\ket{k}\,\sbra{\iota}}{\sbraket{k\,\iota}} \;.
\ee
There are also dressed formulae for the on-shell kinematics of \textit{massive} particles~\cite{Adamo:2020qru}, but since we only consider massless external states they are not needed here. 

\subsection{All-plus amplitudes $({+}{+}{+}{+})$ and collinear limits}

As was the case for QED, a multicollinear configuration of the all-plus helicity, one-loop gluon vacuum amplitude should be obtained from the perturbative expansion of $\cM_{++}$ on a SDPW background. 
In distinction to the abelian setting of QED, we must take into account the colour trace structure of the resulting vacuum amplitude. 
In particular, the perturbative expansion of $\cM_{++}$ will result in \emph{double-trace} gluon amplitudes of a very particular structure: the two initial probe gluons will be in one colour trace with all gluons extracted from the background in the other. 
Since all background gluons are Cartan-valued, the second colour trace will be totally symmetric, meaning that the resulting amplitude is equivalent to a mixed gluon-photon amplitude, with the background gluons effectively serving as photons.

Now, the one-loop, all-positive helicity gluon vacuum amplitude in pure Yang-Mills theory with gauge group SU$(N)$ is given by~\cite{Bern:1993qk,Mahlon:1993si}:
\be\label{glapn}
A^{(1)}_{n}(g_1^+,\ldots,g_n^+)=\frac{\im\,\mathrm{g}^{n}\,N}{\la1\,2\ra\,\la2\,3\ra\cdots\la n\,1\ra}\,\sum_{1\leq i<j<k<l\leq n}\la i\,j\ra\,[j\,k]\,\la k\,l\ra\,[l\,i]\,,
\ee
for the single trace, colour-ordered partial amplitude, where $\mathrm{g}$ is the dimensionless coupling constant and we have dropped an overall numerical factor and suppressed the momentum conserving delta function. 
The double-trace partial amplitudes in this configuration are recovered by using the photon-decoupling identity (cf., \cite{Mangano:1990by,Bern:1990ux}); those double-traces we expect from the perturbative expansion of $\cM_{++}$ are given by
\be\label{dtglapn}
A^{(1)}_{n}(g^+_1,g_2^+|g_3^+,\ldots,g_n^+)=-\!\!\!\sum_{\sigma\in\mathrm{COP}\{2,1\}\{3,\ldots,n\}} A^{(1)}_{n}(g_{\sigma(1)}^+,\ldots,g_{\sigma(n)}^+)\,,
\ee
where $\mathrm{COP}\{2,1\}\{3,\ldots,n\}$ is the set of all permutations of $\{1,\ldots,n\}$ which preserve the cyclic ordering within $\{2,1\}$ and $\{3,\ldots,n\}$ while allowing for all possible relative orderings between the two sets. Identities of this form, combined with the vanishing of all pure-photon amplitudes in the all-plus configuration for $n>4$~\cite{Mahlon:1993fe}, can be used to show that for mixed gluon-photon amplitudes~\cite{Bern:1993qk,Bern:1994ju}
\be\label{mixedvanish}
A^{(1)}_{n}(g_1^+,g_2^+,\gamma_3^+,\ldots,\gamma_n^+)=\left\{
 \begin{array}{lc}
 \im\,\mathrm{g}^4\,N\,\frac{[14]\,[23]}{\la14\ra\,\la23\ra} & \mbox{ if } n=4 \\
 0 & \mbox{ for } n>4
 \end{array}\right.\,.
\ee
Thus, we should find that the perturbative expansion of $\cM_{++}$ on a SDPW background only gives a non-vanishing (double-collinear) result at four-points.

\medskip

The gluon helicity flip amplitude was computed in~\cite{Adamo:2019zmk} for a
general Yang-Mills plane wave background, and is
\begin{align}
\nonumber
\cM_{++} &= \delta^{\mathsf{a}_1\mathsf{a}_2}\,\frac{-\im\,{\mathrm g}^2\,N}{4 \pi^{2}}\,\sum_{e_l} \int\limits_{-\infty}^{+\infty}\!\ud\phi \!\int\limits_0^\infty\!\ud\theta\,\theta\!\int\limits_0^1\!\ud s 
\,\exp\!\bigg[\frac{-\im\theta\,(e_{l} - s e)^2\text{var}(a)}{2 n \cdot k\,s(1-s)}\bigg] (e_{l} - s e)^2 \left(\cA^{2}_{\theta}-\frac{1}{4}\cA^2_{\phi}\right)
\\
\label{ghflip}
&\underset{\text{SDPW}}{\longrightarrow} \delta^{\mathsf{a}_1\mathsf{a}_2}\,\frac{-\im\,\mathrm{g}^2\,N}{4\pi^{2}} \sum_{e_l} \int\limits_{-\infty}^{+\infty}\!\ud\phi \!\int\limits_0^\infty\!\ud\theta\,\theta\!\int\limits_0^1\!\ud s \,(e_{l} - se)^2\bigg(\mathcal{A}_\theta^2 -\frac14 \mathcal{A}^2_\phi \bigg)\,,
\end{align}
where on the second line we have restricted to a SDPW background. Here, the probe gluons come with colour vectors $\mathsf{T}^{\mathsf{a}_1}$, $\mathsf{T}^{\mathsf{a}_2}$, normalised so that $\tr(\mathsf{T}^{\mathsf{a}_1}\mathsf{T}^{\mathsf{a}_2})=\delta^{\mathsf{a}_1\mathsf{a}_2}$, and have charge $e$ with respect to the background. 
The sum is over all colour charges $e_l$ (with respect to the Cartan-valued SDPW background) running in the loop which are compatible with charge conservation.  (For example, with gauge group SU$(2)$ and initial probe charges $e=+1$ with respect to the one-dimensional Cartan subalgebra, only $e_l=0,-1$ are consistent with charge conservation.) 
Structurally, this formula is very similar to the QED helicity flip amplitude: the differences are, aside from overall colour factors, the absence of any exponentials in the SDPW case (as Yang-Mills is massless) and weighting by the colour charge factor $(e_{l} - se)^2$, which introduces new loop momentum dependence.

As \eqref{ghflip} is quadratic in the background, it is immediately obvious that its perturbative expansion will truncate at four-points in vacuum, as required for consistency with \eqref{mixedvanish}. 
The four-point vacuum amplitude is extracted in analogy with the QED calculation: in the perturbative limit the background profile is decomposed into incoming and outgoing Cartan-valued gluons as 
\be\label{ympback}
f(x^\LCm) = \mathrm{g}\,\mathsf{t}\, c_\text{in}\, \e^{-\im\omega\, n\cdot x} + \mathrm{g}\,\mathsf{t}\, c_\text{out}\, \e^{\im\omega\, n\cdot x}\,,
\ee
where $\mathsf{t}$ is a generator of the Cartan subalgebra of the gauge group. 
Upon extracting the coefficient of $c_\text{in}\,c_\text{out}$ in \eqref{ghflip} and changing variables $\theta \to t = \theta/\omega$ one finds
\begin{align}
\mathcal{M}_{++} &= \delta^{\mathsf{a}_1\mathsf{a}_2}\,\frac{-\im\,\mathrm{g}^4\,N}{4\,\pi^2}\,\sum_{e_l}\int\limits_{-\infty}^\infty\!\ud \phi \int\limits_0^1\!\ud s \int\limits_0^\infty\!\ud t\, 
(e_{l} - se)^2\left(\frac{2 \left(t^2-2\right) \cos t-4 t\sin t+4}{t^3}\right)
\frac{{\sbraket{\iota\,k}}^2}{\braket{\iota\,k}^2} \nonumber \\
 &=\delta^{\mathsf{a}_1\mathsf{a}_2}\,\frac{\im\,\mathrm{g}^4\,N}{4\,\pi^2}\,\frac{{\sbraket{\iota\,k}}^2}{\braket{\iota\,k}^2}\,\sum_{e_l}\left(e_l^2-e\,e_l+\frac{e^2}{3}\right)\;, \label{simplifiedym} 
\end{align}
with the symmetric trace over Cartan generators omitted. 
The sum over $e_l$ just contributes a numerical factor to the final answer; its precise value depends on the colour of the initial probe gluons and the rank of the gauge group. (For the SU(2) example above, probe charge $e=+1$ and $e_l=0$ and $-1$, the sum yields $2/3$.) As expected, this is in agreement with the double-collinear limit of the mixed gluon-photon amplitude \eqref{mixedvanish} at four points:
\be\label{ymdcoll}
\lim_{\substack{1,2\to k \\ 3,4\to n}}A^{(1)}_{4}(g_1^+,g_2^+,\gamma_3^+,\gamma_4^+)=\im\,\mathrm{g}^4\,N\,\frac{{\sbraket{\iota\,k}}^2}{\braket{\iota\,k}^2}\,,
\ee
up to an overall numerical factor.

\subsection{One-minus-rest-plus amplitudes and collinear limits}

Again following the analogy with QED, collinear limits of the 1-loop vacuum amplitude with one negative helicity gluon and arbitrarily many positive helicity gluons should be obtained from the perturbative limit of the helicity non-flip amplitude $\cM_{-+}$ on a SDPW background. 
The gluon vacuum amplitude in this helicity configuration is rational and has a compact expression at all multiplicities~\cite{Mahlon:1993si,Bern:2005ji}, but since all gluons extracted from the background are functionally photons, we only need to compare against the mixed gluon/photon amplitudes; there are good reasons to believe that these also obey a vanishing theorem:
\be\label{pmvanish}
A^{(1)}_{n}(g_1^-,g_2^+,\gamma_3^+,\ldots,\gamma_n^+)=\left\{
 \begin{array}{lc}
 \frac{\im\,\mathrm{g}^4}{8\pi^2}\,\frac{\la12\ra\,[24]\,\la41\ra}{\la23\ra\,\la34\ra\,\la24\ra} & \mbox{ if } n=4 \\
 0 & \mbox{ for } n>4
 \end{array}\right.\,.
\ee
Unsurprisingly, the kinematic part of the non-vanishing $n=4$ amplitude is equivalent to the QED result \eqref{4p1n}. 
As a result, the double-collinear limit of the $({-}{+}\cdots{+})$ amplitude vanishes. 
We therefore expect the perturbative limit of $\cM_{-+}$ to vanish as well.

In fact, gluon helicity non-flip on a general Yang-Mills plane wave
background has not been calculated previously, so we provide details of the
calculation in Appendix~\ref{app:YMNonFlip}; the final result is
\begin{align}\label{YMnf*}
    \mathcal{M}_{-+}
    =
    -
    \im
    &
    \delta^{\mathsf{a}_{1}\mathsf{a}_{2}}
    \frac{
        \mathrm{g}^{2}
        N
    }{8 \pi^{2}} 
    \int_{-\infty}^{\infty} \ud \phi
    \int_{0}^{\infty} \ud \theta \;
    \theta
    \int_{0}^{1} \ud s \;
    \e^{
        -
        \im
        \frac{\theta \text{var}(a)\, (e_{l} - s e)^{2}}{2 n \cdot k\, s (1 - s)}
    }
    (e_{l} - se)^{2}
    \nonumber\\
    &
    \times
    \bigg[
        \epsilon^{(+)} \cdot \epsilon^{(-)}
        \bigg(
            \frac{1}{\theta^{2}}
            \partial_{\theta}
            \Big[\theta \text{var}(a)\Big]
            -
            \frac{\widehat{a}_{\phi}\cdot\widehat{a}_{\phi}}{(1-s) s}
            \bigg(
                1
                -
                \frac{1}{2(1-s)s}
            \bigg)
        \bigg)
        \nonumber\\
        &
        \qquad
        +
        2
        \cA_{\theta} \bar{\cA}_{\theta}
        -
        \frac{1}{2}
        \cA_{\phi} \bar{\cA}_{\phi}
        +
        \bigg(
            1
            -
            \frac{2}{(1-s)s}
        \bigg) 
        \bar{\cA}_{[\phi}
        \cA_{\theta]}
    \bigg]
\end{align}
where the gauge group is SU$(N)$. 
As in QED, the first term proportional to $\epsilon^{(+)} \cdot \epsilon^{(-)}$ has been regularised by subtracting off the vacuum contribution, and so is UV finite.
Comparing with the corresponding QED amplitude~\eqref{simplify-me}, the general structure here is similar, though there are some notable differences. 
In particular, there is a different dependence on the lightfront momentum fraction $s$, which goes beyond the weighting by colour charge factors. This is in contrast to the comparison of the photonic and gluonic all-plus amplitudes \eqref{simplified} and \eqref{ghflip}.

It is easy to see that $\cM_{-+}$ vanishes for a SDPW background, as in that case $\mathrm{var}(a)$, $\bar{\cA}$ and $a^2$ all vanish, and hence the integrand of \eqref{YMnf*} vanishes. 
$\cM_{-+}$ is therefore zero on the self-dual background, in accordance with the double collinear limits of the conjectured vanishing theorem \eqref{pmvanish}.

\subsection{QCD}

It is fairly straightforward to extend these results to QCD with $n_f$ flavours of fundamental-valued quarks, even with distinct masses, as the additional contributions from the quarks closely mimic those of the electron in QED. 
On the vacuum amplitude side, the vanishing theorems \eqref{mixedvanish} and \eqref{pmvanish} still hold with quark contributions inside the loop; once again, only the $({+}{+}{+}{+})$ amplitude will survive in the double collinear limit. 
The relevant mixed gluon/photon amplitude for $n=4$ is given by simply including the quark loop contributions~\cite{Bern:1995db}:
\be\label{QCD1}
-\im\,\mathrm{g}^4\,\frac{[14]\,[23]}{\la14\ra\,\la23\ra}\sum_{i=1}^{n_f}\left(m_i^{4}\,I_{4}-\frac{1}{6}\right)\,,
\ee
up to an overall numerical factor, where $I_4$ is a scalar box integral and $\{m_1,\ldots,m_{n_f}\}$ are the quark masses (see Appendix~\ref{app:MASS} for details). 

The contribution of quarks to $\cM_{++}$ was computed in~\cite{Adamo:2019zmk} for a general plane wave background:
\begin{multline}\label{quarkloop}
\cM^{\mathrm{quark}}_{++} = \delta^{\mathsf{a}_1\mathsf{a}_2}\,\frac{\im\,\mathrm{g}^2}{(2\pi)^{2}}\,\sum_{i=1}^{n_f}\sum_{\mu_l} \int\limits_{-\infty}^{+\infty}\!\ud\phi \!\int\limits_0^\infty\!\ud\theta\,\theta\!\int\limits_0^1\!\ud s 
\,\exp\!\bigg[- \im \frac{\theta\,M_{i}^{2}[\mu_l]}{2 n \cdot k\, s(1-s)}
\bigg] \\
	 \times (\mu_l - s e)^2 \left(\cA^{2}_{\theta}-\frac{1}{4}\cA^2_{\phi}\right)\,,
\end{multline}
where the second summation is over the fundamental charges $\mu_l$ of the quarks in the loop with respect to the Cartan-valued background and $M_{i}^{2}[\mu_l]$ is the Kibble mass of each quark
\be\label{qKM}
M_{i}^{2}[\mu_l]:=m^2+(\mu_l-se)^2\,\mathrm{var}(a)\,.
\ee
Restricting to a SDPW background is straightforward, and combining with the contribution from pure Yang-Mills, the perturbative limit yields:
\begin{multline}\label{QCD2}
\cM^{\mathrm{QCD}}_{++}\to \delta^{\mathsf{a}_1\mathsf{a}_2}\,\frac{\im\,\mathrm{g}^4}{4\,\pi^2}\,\frac{{\sbraket{\iota\,k}}^2}{\braket{\iota\,k}^2}\left[N\sum_{e_l}\left(e_l^2-e\,e_l+\frac{e^2}{3}\right)\right. \\
\left.+\sum_{i=1}^{n_f}\sum_{\mu_l} \int\limits_{-\infty}^\infty\!\ud \phi \int_0^1\!\ud s \int\limits_0^\infty\!\ud t\, 
\e^{\frac{-\im\,t\, m_i^2}{2\omega\, n\cdot k\, s(1-s)}}\,(\mu_l - s e)^2
\left(\frac{2 \left(t^2-2\right) \cos t-4 t\sin t+4}{t^3}\right)\right]\,.
\end{multline}
For the simplest case of $n_f$ \textit{massless} quarks, all integrals are readily evaluated to give 
\be\label{QCD} 
\delta^{\mathsf{a}_1\mathsf{a}_2}\,\frac{\im\,\mathrm{g}^4}{4\,\pi^2}\,\frac{{\sbraket{\iota\,k}}^2}{\braket{\iota\,k}^2}\,\left[N\sum_{e_l}\left(e_l^2-e\,e_l+\frac{e^2}{3}\right) - n_f \sum_{\mu_l}\left(\mu_l^2-e\,\mu_l+\frac{e^2}{3}\right) \right]\;,
\ee
which generalises the pure Yang-Mills results \eqref{simplifiedym} and is clearly consistent with the double collinear limit of \eqref{QCD1} in the massless case. 
Correspondence with the double collinear limit of \eqref{QCD1} for non-zero quark mass is established using the same methods as employed for massive QED in Appendix~\ref{app:MASS}.


%% file: SUSY.tex
%
The low-energy (photonic) dynamics of QED is described by the Euler-Heisenberg effective Lagrangian which ascribes, to constant or slowly varying electromagnetic fields, two vacuum refractive indices. 
These modify the propagation of probe photons, leading to the effect of vacuum birefringence~\cite{Heinzl:2006xc}; that is, the polarisation rotation of a beam of photons traversing a strong electromagnetic background (clearly a genuine loop effect; see~\cite{King:2015tba,Sangal:2021qeg,Gies:2021ymf} for recent and related investigations). 
This effect is supported on the difference of the refractive indices, but in supersymmetric QED (as in e.g.~Born-Infeld theory), the refractive indices are equal~\cite{Duff:1979bk}, so there is no vacuum birefringence~\cite{Rebhan:2017zdx} (see also \cite{Cheung:2018oki}).

In terms of scattering amplitudes, though, and beyond the Euler-Heisenberg limit, vacuum birefringence is really the macroscopic result of a non-vanishing photon helicity flip amplitude, that is $\cM_{++}$~\cite{Dinu:2013gaa}. 
As such, we here extend the preceding calculations from QED to $\mathcal{N}=1$ supersymmetric QED (SQED), and from Yang-Mills to $\cN=1$ super-Yang-Mills (SYM). Note that we will consider completely \emph{general} (not just self-dual) plane wave backgrounds.

\medskip

To calculate the helicity flip amplitude in SQED we need to add, to the QED amplitude, the contribution of two additional scalars running in the loop. 
Both the QED and scalar QED helicity flip amplitudes on a general plane wave background may be found together in~\cite{Ilderton:2016qpj}; the scalar result is given by exactly $-1/2$ times the fermion result \eqref{simplified}:
\be
	\mathcal{M}^\text{scalar}_{++} = -\frac{1}{2}\,
	\mathcal{M}_{++} \;.
\ee
So in SQED, where one fermion and two scalars run in the loop, the contributions cancel and the flip amplitude vanishes ($\cM_{++}^{\mathrm{SQED}}=0$) on \emph{any} plane wave background, at \emph{all} energies. 

\medskip

To establish whether a similar result exists in \emph{non-abelian} supersymmetric gauge theories, we consider $\cN=1$ super-Yang-Mills (SYM) theory. 
The only new ingredient in the gluon helicity flip amplitude for this theory (as compared to pure Yang-Mills) is the presence of a gluino loop. 
Recall that the gluino, the super-partner of the gluon, is a massless Weyl fermion valued in the adjoint representation of the gauge group. 
The calculation of its contribution to gluon helicity flip requires only the gluino propagator, defined by the kinetic term in the $\cN=1$ SYM Lagrangian
\be\label{gluinoKT}
\mathcal{L}^{\mathrm{gluino}}_{\mathrm{kin}}=\tr\left(\tilde{\Psi}^{\dot\alpha}\,D_{\alpha\dot\alpha}\Psi^{\alpha}\right)\,,
\ee
where $D_{\alpha\dot\alpha}$ is the gauge connection of the plane wave background. 
This is easily constructed using ingredients encountered above and in~\cite{Adamo:2019zmk}. 

The calculation contains no surprises, so we present instead a simple argument to obtain the result directly from the quark loop term for helicity flip in QCD, given by \eqref{quarkloop}. 
Let the $n_f$ quarks, valued in the fundamental representation of the gauge group, have identical masses. 
To obtain the analogous contribution for \emph{adjoint}-valued fermions, one simply replaces $\mu_{l}\rightarrow e_{l}$, and $n_f$ with $2N$, where the factor of two arises from the distinction between the Casimir invariants $T_{\mathrm{fund}}=\frac{1}{2}$ and $T_{\mathrm{adj}}=1$. 
To account for the distinction between Dirac and Weyl/Majorana fermions the result must be multiplied by a factor of $\frac{1}{2}$. Finally, setting the mass to zero yields
\begin{multline}\label{gluinoloop}
\cM^{\mathrm{gluino}}_{++}= \delta^{\mathsf{a}_1\mathsf{a}_2}\,\frac{\im\,g^2\,N}{(2\pi)^{2}}\,\sum_{e_l} \int\limits_{-\infty}^{+\infty}\!\ud\phi \!\int\limits_0^\infty\!\ud\theta\,\theta\!\int\limits_0^1\!\ud s 
\,\exp\!\bigg[- \im \frac{\theta\,\text{var}(a)}{2 n \cdot k\,s(1-s)}
\,(e_{l} - s e)^2\bigg] \\
	\times (e_{l} - s e)^2 \left(\cA^{2}_{\theta}-\frac{1}{4}\cA^2_{\phi}\right)\,.
\end{multline}
Comparing to the Yang-Mills result~\eqref{ghflip}, we see immediately that $\cM^{\mathrm{gluino}}_{++}=-\cM^{\mathrm{YM}}_{++}$; the gluino loop exactly cancels the pure Yang-Mills contributions and thus $\cM^{\mathrm{SYM}}_{++}=0$ in $\cN=1$ SYM -- the gluon helicity flip amplitude vanishes.

\medskip

This gives an explicit generalisation of existing results beyond the Euler Heisenberg limit (i.e.~to arbitrary energies) in both abelian and non-abelian settings, as well as to arbitrary plane wave backgrounds with any strength and profile. 
The phenomenological consequence is that in SUSY theories, there is no vacuum birefringence in \textit{any} plane wave background.

\medskip

The reason behind this is not immediately obvious. 
In particular, it is not clear that $\cM_{++}=0$ as a consequence of any SUSY Ward identity, since the purely photonic/gluonic plane wave is not a fully supersymmetric solution. 
Recall, though, that a key component of helicity conservation in $2\to 2$ SUSY processes~\cite{Gounaris:2005ey,Gounaris:2006zm} is that the SUSY transform between gluons and gluinos, when projected onto \textit{physical} one-particle states, is helicity conserving. 
We conjecture that this property extends to scattering on plane-wave backgrounds, and is therefore responsible for the vanishing of helicity flip. 

To support this, note that propagation through the plane wave conserves individual particle helicity~\cite{Ilderton:2020gno}, despite the fact that the plane wave is not a SUSY background. 
This is already clear from the dressed gluon polarisation vectors \eqref{GLUON-POL-DRESSED}, which are obtained from their vacuum counterparts simply by replacing the asymptotic particle momentum with the dressed momentum. Conservation of helicity can also be made explicit for fermions, as dressed spinors $U$ are related to undressed spinors $u$ by  
               \be
               		U = \bigg(1\!\!1 + \frac{\slashed{n} \slashed{a}(x^\LCm)}{2n\cdot p}\bigg)u  = \bigg(1\!\!1  + \frac{{\bar f}(x^\LCm) \ket{\iota}\bra{\iota}- f(x^\LCm) \sket{\iota}\sbra{\iota}}{\sqrt{2} n\cdot p} \bigg) u\;,
               \ee
so what is initially left (right)-handed remains left (right)-handed after entering the sandwich plane wave. Further, only the transversely projected (i.e.,~physical) helicity states in the loop contribute to helicity flip~\cite{Dinu:2013gaa,Adamo:2019zmk}. 
Put together, these results suggest that the helicity conservation of SUSY theories continues to hold in general plane wave backgrounds; it would be interesting to investigate this in more detail.

%% file: Conclusions.tex
%

In this paper, we extracted maximally collinear limits of one-loop scattering amplitudes in vacuum through the expansion of one-loop, $2$-point amplitudes in gauge theories on self-dual plane wave backgrounds. These 2-point amplitudes are calculated for arbitrary background strength and profile due to the high degree of symmetry of the plane wave, and as such contain structures not found in vacuum amplitudes, which in turn leads to novel physics such as vacuum birefringence. Restricting to self-dual backgrounds yields some significant simplifications and gives direct access to well-studied helicity configurations, in particular all-plus and one-minus-rest-plus. While self-duality of the background ensures that the perturbative expansion truncates at 4-point vacuum amplitudes, matching the perturbative expressions with those vacuum amplitudes is surprisingly subtle, particularly in the massive case and at higher orders (see appendices \ref{app:MASS}, \ref{app:six-point}).


Despite the complexity of the amplitudes themselves, we have demonstrated that, in general plane wave backgrounds, the helicity flip amplitude vanishes in $\cN=1$ supersymmetric QED and Yang-Mills theory. This result was already known in the low-energy (Euler-Heisenberg) limit of QED, but our results hold for non-abelian gauge theories, arbitrary energies and non-constant plane wave backgrounds of arbitrary strength.
        
\medskip
        
The results presented here could be extended in a number of ways. The double copy relationship between vacuum scattering amplitudes in Yang-Mills theory and gravity has been shown to hold to high-multiplicity and loop level (see \cite{Bern:2019prr} for a review). It has also been extended to low-multiplicity tree-level scattering amplitudes on plane wave background fields~\cite{Adamo:2017nia,Adamo:2017sze,Adamo:2018mpq,Adamo:2020qru}, giving access to non-perturbative features of scattering on gravitational waves. Generalising the results of the present paper to the background field double copy would be a natural step in further exploring this relationship, and the collinear aspects of the corresponding vacuum gravitational amplitudes at one loop.

One could also consider higher-point, and higher-loop order amplitudes on the background. For example -- and unlike in vacuum -- the $3$-point photon amplitude on a background is nonvanishing (see e.g.~\cite{DiPiazza:2007yx} for the plane wave case) as Furry's theorem does not apply. The perturbative expansion of this amplitude would give direct access to helicity configurations beyond those considered here.
        

%% file: massive.tex
In Sect.~\ref{sec:QED} we demonstrated that by perturbatively expanding $2$-point amplitudes on a plane wave background we could reproduce the maximally collinear limits of scattering amplitudes in vacuum, in massless QED. We show here that the results also hold for massive QED, limiting our discussion to the all-plus helicity configuration as only this one can be non-vanishing on a SDPW background, as is clear from \eqref{simplified}) and \eqref{simplify-me}.

The all-plus scattering amplitude in massive QED is given by (see e.g.~\cite{Bern:1995db}),
\begin{align}\label{eqn:MassiveBern}
    A^{(1)}_{4}(\gamma_1^{+},\gamma_2^{+},\gamma_3^{+},\gamma_4^{+})
    =
    \sum_\text{perms}
    \frac{\im}{16 \pi^{2}}
    \frac{{\sbraket{12}}\,{\sbraket{34}}}{\braket{12}\,\braket{34}}
    \Big(
        m^{4}\,
        I_{4}
        -
        \frac{1}{6}
    \Big) \,,
\end{align}
where the sum is over permutations of $k_1 \ldots k_4$ and the scalar box integral $I_{4}$ is defined by 
\be\label{I4def}
I_{4}
=
\int\!\frac{\ud^4p}{(2\pi)^4} \, \frac{-16\,\im\, \pi^2 }{(p^2-m^2) \big( (p-k_1)^2-m^2\big) \big( (p-k_1-k_2)^2-m^2\big)\big( (p+k_4)^2-m^2\big)} \;.
\ee
The spinor structure is actually independent of the ordering, so can be brought outside the sum.  
Performing the momentum integral in \eqref{I4def}, one obtains an expression in terms of the dilogarithm; our expression for the massive amplitude is, however, \eqref{simplified4} and the $s$-integral can be expressed in terms of Bessel functions using
	\be
		\int\limits_0^1\!\ud s \, \e^{-\im \frac{x}{2s(1-s)}} = \im\, x\, \e^{-\im x} \big ( K_1(\im x) - K_0(\im x)\big) \;.
	\ee
To show equivalence between \eqref{I4def} and \eqref{simplified4}, one must convert the scalar box $I_4$ back to position space, and evaluate the integrals in lightfront coordinates, in order to match the calculation on the SDPW background.
 
The $p_\LCm$ integral of the scalar propagator is evaluated by residues first, yielding the lightfront form of the propagator
\be
	G(x) = \im \int\!\frac{\ud^4 p}{(2\pi)^4} \frac{\e^{-\im p\cdot x}}{p^2-m^2 + \im \epsilon} =  \int\!\frac{\ud^2 p_\LCperp\, \ud p_\LCp}{(2\pi)^3\,|2p_\LCp|} \,\Theta(p_\LCp x^\LCm)\, \e^{-\im p\cdot x} \;.
\ee
We then integrate out the `spatial' coordinates~$x^\LCperp$ and~$x^\LCp$, at each vertex. The two $x^\LCm$ integrals are evaluated against the various step functions, and give trigonometric functions of the loop momentum $s$, as in \eqref{simplified4}. Changing to the average ($\phi$) and difference ($\theta \equiv t/\omega$) phase variables for the two remaining $x^\LCm$ integrals, one finds
	\be\label{nearly1}
		\sum_\text{perms} \Big(m^4 I_4 - \frac{1}{6}\Big) = -1 - \frac{2}{b^2} \int\limits_0^\infty\! \ud t \int\limits_0^1\!\ud s \,  \e^{-\frac{\im}{2b} \frac{t}{s(1-s)}} \bigg[ \frac{1}{(1-s)^2} + \frac{1}{s(1-s)}\bigg]  \frac{\sin^2(t/2)}{t} \;,
	\ee
in which $b:= \omega k^\LCp/m^2$. 

Despite appearances, this is indeed the same as the massive QED expression \eqref{simplified4}. The key lies in the first term in square brackets of \eqref{nearly1} which, as it is not a function of $s(1-s)$, appears be to be incompatible with \eqref{simplified4}. However, the exponential which appears in all our expressions is symmetric in $s\leftrightarrow 1-s$ over the $s$-integration range, meaning that the \textit{pre}-exponential can be symmetrised; the terms in square brackets of \eqref{nearly1} then become
	\be
	\frac{1}{2}\bigg(\frac{1}{(1-s)^2} + \frac{1}{s^2}\bigg) + \frac{1}{s(1-s)} = \frac{1}{2}\frac{1}{s^2(1-s)^2} \;.
	\ee
This is, up to factors, just the second derivative of the exponent with respect to $t$. Writing the integrand as such, we integrate by parts in $t$. There is no boundary contribution from the first integration, while the second yields a nonzero, finite contribution at $t=0$; this boundary term exactly cancels the `rational part' subtracted from $I_4$. Evaluating the derivatives in the remaining bulk term gives
	\be\label{nearly2new}
	\begin{split}
		\sum_\text{perms} \Big(m^4 I_4 - \frac{1}{6}\Big) 
 = \int\limits_0^\infty\! \ud t \int\limits_0^1\!\ud s \,  \e^{-\frac{\im}{2b} \frac{t}{s(1-s)}} \frac{2(t^2-2)\cos t-4 t \sin t + 4}{t^3} \;,
	\end{split}
	\ee
exactly reproducing our expression \eqref{simplified4} for the massive QED result.
	
%
%

%% file: appendix-six-point.tex

We have seen that the restriction to a SDPW background ensures that the perturbative expansion of both helicity flip and non-flip amplitudes cannot result in higher than 4-point vacuum amplitudes.
%
%
However, on general, \emph{non-chiral} plane wave backgrounds, it is possible to obtain non-vanishing vacuum amplitudes in these helicity configurations at higher points, since there is an additional, \textit{exponential} dependence on the background entering through the quantity $\mathrm{var}(a)$ defined by \eqref{Kibblem}, and each insertion of $\mathrm{var}(a)$ obtained by Taylor-expanding the exponential will lead to a pair of external photons of opposite helicity. 

For example, expanding the helicity flip amplitude $\cM_{++}$ of \eqref{simplified} to first order in the exponent generates a (nonzero) six-photon vacuum amplitude in the helicity configuration $(+++++-)$. 
As long as $m\neq0$, there is no contradiction posed by this expression, since the 6-point, 1-loop vacuum amplitude with one (or zero) negative helicity photons is non-vanishing in \textit{massive} QED (cf., \cite{Nagy:2006xy,Ossola:2007bb,Bernicot:2008th}). 
However, in the massless case this expression must vanish to be consistent with Mahlon's vanishing theorem~\cite{Mahlon:1993fe}\footnote{For {non-vanishing} helicity configurations at 6-points see e.g.~\cite{Binoth:2007ca}}. 
Setting $m=0$, one obtains the six-photon expansion of the helicity-flip amplitude:
\be\label{expanded-6pt}
	\mathcal{M}_{++} \to -\frac{\alpha}{n\cdot k\, \pi} 
\int\limits_0^1\! \frac{\ud s}{s(1-s)}
\int\limits_{-\infty}^\infty\!\ud \phi \int\limits_0^\infty\!\ud\theta\,\theta^2 \bigg(\mathcal{A}_\theta^2 -\frac14 \mathcal{A}^2_\phi \bigg)\,  \text{var}(a)\;,
\ee 
where two positive helicity photons come from the probe on the background, two more positive helicity photons come from the pre-exponential, and a positive/negative helicity pair of photons is encoded in $\mathrm{var}(a)$. 

The $s$-integral in \eqref{expanded-6pt} is divergent in massless QED and requires regularization. 
One can either insert explicit cutoffs into the $s$--integral, or perform the entire calculation in transverse dimensional regularization, going to $2+2\varepsilon$ transverse directions (see~\cite{Casher:1976ae} for details). We will work with the latter method, which has the benefit of not changing any properties of the background; this results in an additional factor of $(s(1-s))^{\varepsilon}$ in \eqref{expanded-6pt}. The $s$-integral is then convergent for $\varepsilon>0$, with a simple expression in terms of gamma functions, so we leave the regularization implicit from now on.

Now, take the background dependence in \eqref{expanded-6pt} to be the momentum eigenstate version of \eqref{gpw}:
	\be
		a_\mu = \e^{\im \lambda x^\LCm}(0,0,1,\im) + \e^{-\im\omega x^\LCm}(0,0,1,-\im) \;,
	\ee
where we ignore overall normalizations for the time-being. 
In terms of physical variables, with $\lambda,\omega>0$ this field describes incoming negative helicity photons of energy $\omega$ and outgoing negative helicity photons of energy $\lambda$. 
The important building blocks of the amplitude are easily evaluated; for example,
	\be
		\mathcal{A} = - \e^{\im\lambda\phi} \,\text{sinc}\!\left(\frac{\lambda\theta}{2}\right)  \;, 
	\ee
while $\text{var}(a)$ has the somewhat unrevealing expression	
\begin{multline}\label{pvar}
\text{var}(a) = \frac{\e^{\im(\lambda -\omega ) \phi  }}{\lambda-\omega}\left[\mathrm{sinc}\!\left(\frac{\theta  \lambda }{2}\right) \left(\lambda\,\cos\!\left(\frac{\theta  \omega }{2}\right)+(\omega -\lambda )\, \mathrm{sinc}\!\left(\frac{\theta  \omega }{2}\right)\right) \right. \\
\left.-\omega\,  \cos\!\left(\frac{\theta  \lambda }{2}\right)\, \mathrm{sinc}\!\left(\frac{\theta  \omega }{2}\right)\right]\,,
\end{multline}
Observe that the $\phi$-dependence of $\text{var}(a)$ is $\exp \im(\lambda-\omega)\phi$, and that the product of (derivatives of) $\mathcal{A}$ go like $\exp 2\im\lambda \phi$. 
Thus, the $\phi$-integral in \eqref{expanded-6pt} yields the `missing' overall conservation of momentum expected for vacuum amplitudes, fixing $3\lambda = \omega$. 

What remains, upon changing variables $\theta \to t := \lambda \theta$, is given by
\be\label{expanded-6pt2}
 \delta(3\lambda - \omega)\, \int\limits_0^1\! \frac{\ud s}{s(1-s)}
\int\limits_0^\infty\!\frac{\ud t}{t^4}\, \sin\!\left(\frac{t}{2}\right) \left(t^2-2 t \sin (t)-2 \cos (t)+2\right) \left[3 t \cos\!\left(\frac{t}{2}\right)-2 \sin\!\left(\frac{3 t}{2}\right)\right]\,,
\ee
up to a constant of proportionality. 
While cumbersome, the $t$-integral is elementary and evaluates to zero. 
Since the remaining $s$-integral is regularized, this kills the entire amplitude, as desired. 
Hence, we see that higher-order perturbative expansions on non-chiral plane wave backgrounds remain consistent with the vanishing theorems of massless QED at one-loop.

%% file: Appendix_No_Flip_YM.tex

The gluon helicity flip amplitude for Yang-Mills on a general plane wave background is given in~\cite{Adamo:2019zmk}, but the corresponding non-flip amplitude has not previously appeared in the literature. Its calculation proceeds as for helicity flip~\cite{Adamo:2019zmk}, to which we refer the reader for details. Here we present an outline, working in $d = 4$.

We consider the (necessarily forward, see below) scattering of an on-shell gluon with initial momentum $k_{\mu}$ and polarisation $\epsilon_{\mu}$, where $k^{2} = 0 = k \cdot \epsilon$, and work in lightcone gauge $n \cdot \varepsilon = 0$. 
Before entering the sandwich wave background, the gluon is described by the wavefunction $\mathsf{T}^{\mathsf{a}} \epsilon_{\mu} \e^{-\im k \cdot x}$, where $\mathsf{T}^{\mathsf{a}}$ is the generator of the gauge group and $\mathsf{a} = 1,\dots,N^{2} - 1$ for SU$(N)$. 
Inside the background field the polarisation and momentum of the gluon become dressed, such that the wavefunction is
\begin{align}\label{eqn:DressedGluon}
    \mathsf{T}^{\mathsf{a}}
    \mathcal{E}_{\mu}(x^{\LCm})
    \exp\big[- \im \phi_{k}(x)\big] 
    \;, \quad 
    \text{where}
    \quad
    \phi_{k}(x)
    =
    k \cdot x
    +
    \int_{-\infty}^{x^{\LCm}}
    \ud t\,
    \frac{2 e a(t) \cdot k - e^{2} a^{2}(t)}{2 n \cdot k}
    \,,
\end{align}
and $e$ is the (Cartan) colour charge of the initial gluon (as described below \eqref{glumom}).
The dressed momentum of the gluon, $K_{\mu}(x^{\LCm})$, is defined by
\begin{align}\label{eqn:DressedMomentum}
    K_{\mu}(x^{\LCm})   
    =
    \bigg(
        k_{\mu}
        -
        e a_{\mu}(x^{\LCm})
        +
        \frac{2 e a(x^{\LCm}) \cdot k - e^{2}a^{2}(x^{\LCm})}{2 n \cdot k}
        n_{\mu}
    \bigg) 
    \;,
\end{align}
and is obtained through acting on the exponential with the background field covariant derivative, $ K_{\mu}(x^{\LCm}) = \im \e^{\im\phi_{k}(x)} D_{\mu} \e^{-\im\phi_{k}(x)}$. 
Alternatively, it can be written as a field and momentum-dependent Lorentz boost acting on the asymptotic momentum $k_{\mu}$,
\begin{align}\label{eqn:Boost}
    K_{\mu}(x^{\LCm})
    =
    \bigg(
        \eta_{\mu\nu}
        -
        \frac{
            e a_{\mu}(x^{\LCm}) n_{\nu}
            -
            e a_{\nu}(x^{\LCm}) n_{\mu}
        }{n \cdot k}
        -
        \frac{e^{2} a^{2}(x^{\LCm}) n_{\mu} n_{\nu}}{2 (n \cdot k)^{2}}
    \bigg)
    k^{\nu}
    \equiv
    \Lambda_{\mu\nu}(x^{\LCm};k)
    k^{\nu}
    \,,
\end{align}
which also defines the dressed polarisation
\begin{align}\label{eqn:DressedPolarisation}
    \mathcal{E}_{\mu}(x^{\LCm})
    =
    \Lambda_{\mu\nu}(x^{\LCm};k)
    \epsilon^{\nu}
    \,.
\end{align}

Working in Feynman-'t Hooft gauge, we need to calculate both the gluon loop and ghost loop contributions to the amplitude.  
The background dressed gluon propagator is
\begin{align}\label{eqn:GluonPropagator}
    \mathcal{G}_{\mu\nu}^{\mathsf{a}_{1}\mathsf{a}_{2}}(x,y)
    =
    -
    \im \delta^{\mathsf{a}_{1}\mathsf{a}_{2}}\,
    \oint
    \frac{\ud^{4} l}{(2\pi)^{4}}
    \frac{
        D^{l}_{\mu\nu}(x,y)
    }{l^{2}+\im\varepsilon}
    \e^{-\im\phi_{l}(x) + \im\phi_{l}(y)}
    \,,
\end{align}
where the tensor structure is, using \eqref{eqn:Boost},
\begin{align}\label{eqn:PropagatorStructure}
    D^{l}_{\mu\nu}(x,y)
    =
    \Lambda_{\mu\lambda}(x^{\LCm};l)\Lambda_{\nu}^{~\lambda}(y^{\LCm};l)
    \,,
\end{align}
with $l_{\mu}$ continued off-shell.
The corresponding ghost propagator is just \eqref{eqn:GluonPropagator} with the tensor structure set to unity.
For the background field $2$-point amplitude we require only the $3$-point gluon and gluon-ghost vertices. 
The gluon tadpole contributes only background-independent terms equal to those in vacuum, and so is subtracted entirely by renormalization, see~\cite{Adamo:2019zmk} for details. 
These vertices are, respectively,
\begin{align}\label{eqn:GluonVertex}
    \Gamma^{\mathsf{a}_{1}\mathsf{a}_{2}\mathsf{a}_{3}}_{\mu\nu\sigma}(x)
    &=
    \mathrm{g} 
    f^{\mathsf{a}_{1}\mathsf{a}_{2}\mathsf{a}_{3}}
    \int \ud^{4}x
    \Big[
        \eta_{\mu\nu} \big(D_{1} - D_{2}\big)_{\sigma}
        +
        \eta_{\nu\sigma} \big(D_{2} - D_{3}\big)_{\mu}
        +
        \eta_{\sigma\mu} \big(D_{3} - D_{1}\big)_{\nu}
    \Big] 
    \,,
    \\
\label{eqn:GhostVertex}
    \Gamma^{\mathsf{a}_{1}\mathsf{a}_{2}\mathsf{a}_{3}}_{\mu}(x)
    &=
    \mathrm{g} 
    f^{\mathsf{a}_{1}\mathsf{a}_{2}\mathsf{a}_{3}}
    \int \ud^{4}x\,
    D_{1\mu}
    \,,
\end{align}
where $f^{\mathsf{a}_{1}\mathsf{a}_{2}\mathsf{a}_{3}}$ are the colour structure constants, and a subscript on the background covariant derivative denotes the particle on which it acts. 

With these ingredients one can write down the no-flip amplitude and proceed by evaluating position and momentum integrals. 
Due to the non-trivial space-time dependence of the background only the integrations over $x^{+}$ and $x^{\LCperp}$ can be performed at each vertex to give (six) momentum-conserving $\delta$-functions, which eliminate three loop momenta. 
Two further loop integrals in $l_{\LCm}$ and $l^{\prime}_{\LCm}$, for propagator momentum $l_\mu$ and $l^{\prime}_{\mu}$, respectively, are performed by contour integration, putting them on-shell.
Written covariantly, the momentum $l^{\prime}$ is then
\begin{align}\label{eqn:Covariantlp}
    l^{\prime}_{\mu}
    =
    l_{\mu}
    -
    k_{\mu}
    +
    \frac{(k - l)^{2}}{2 n \cdot (k - l)}\,
    n_{\mu}
    \,,
\end{align}
and it can easily be shown that this relationship extends to the dressed momenta ($L$ for dressed $\ell$, $K$ for dressed $k$, etc.)
\begin{align}\label{eqn:DressedLp}
    L^{\prime}_{\mu}(x^{\LCm})
    =
    L_{\mu}(x^{\LCm})
    -
    K_{\mu}(x^{\LCm})
    +
    \frac{(K(x^{\LCm}) - L(x^{\LCm}))^{2}}{2 n \cdot (k - l)}\,
    n_{\mu}
    \,.
\end{align}
The remaining loop integrals are over $l_\LCperp$ and $s = n \cdot l / n \cdot k$.

The amplitude then takes the form
\begin{align}\label{eqn:Amplitude}
    \mathcal{A}
    =
    (2\pi)^{3} \delta_{\LCp,\LCperp}^{3}(k^{\prime} + k) \mathcal{M}(k)
    \,,
\end{align}
with
\begin{align}\label{eqn:GhostAmp1}
    \mathcal{M}(k)
    =
    \frac{
        \mathrm{g}^{2}
        N\,
        \delta^{\mathsf{a}_{1}\mathsf{a}_{2}}
    }{8 k_{\LCp} (2\pi)^{3}} 
    \int \ud \phi
    \int_{0}^{\infty} \ud \theta
    \int_{0}^{1} 
    \frac{\ud s}{s (1 - s)}
    &
    \int \ud^{2} l_{\LCperp}
    \exp\bigg[
        \im
        \int_{\phi - \theta/2}^{\phi + \theta/2} \ud t\,
        \frac{K(t) \cdot L(t)}{n \cdot k (1 - s)}
    \bigg] 
    \nonumber\\
    &
    \times
    \bigg(
        \mathcal{T}_{\text{ghost}}
        -
        \frac{1}{2}
        \mathcal{T}_{\text{gluon}}
    \bigg)
    \,,
\end{align}
for the tensor structures
\begin{align}\label{eqn:GhostTensor}
    \mathcal{T}_{\text{ghost}}
    =
    \mcE(x^{\LCm}) \cdot L(x^{\LCm})
    \mcE^{\prime}(y^{\LCm}) \cdot L(y^{\LCm})
    \,,
\end{align}
and
\begin{align}\label{eqn:GluonTensor}
    \mathcal{T}_{\text{gluon}}
    =
    &
    \Big[
        -
        \eta^{\mu\rho}
        \big(
            K^{\sigma}(x^{\LCm})
            +
            L^{\sigma}(x^{\LCm})
        \big)
        +
        \eta^{\rho\sigma}
        \big(
            L^{\mu}(x^{\LCm})
            +
            {L^{\prime}}^{\mu}(x^{\LCm})
        \big)
        +
        \eta^{\sigma\mu}
        \big(
            K^{\rho}(x^{\LCm})
            -
            {L^{\prime}}^{\rho}(x^{\LCm})
        \big)
    \Big] 
    \nonumber\\
    &
    \times
    \mcE_{\mu}(x^{\LCm})
    \mcE^{\prime}_{\nu}(y^{\LCm})
    D^{l - k}_{\sigma\alpha}(x^{\LCm},y^{\LCm})
    D^{l}_{\beta\rho}(y^{\LCm},x^{\LCm})
    \nonumber\\
    &
    \times
    \Big[
        \eta^{\nu\alpha}
        \big(
            {K}^{\beta}(y^{\LCm})
            - 
            {L^{\prime}}^{\beta}(y^{\LCm})
        \big)
        +
        \eta^{\alpha\beta}
        \big(
            {L^{\prime}}^{\nu}(y^{\LCm})
            +
            {L}^{\nu}(y^{\LCm})
        \big)
        -
        \eta^{\beta\nu}
        \big(
            {L}^{\alpha}(y^{\LCm})
            +
            {K}^{\alpha}(y^{\LCm})
        \big)
    \Big] 
    \,.
\end{align}
The exponent in \eqref{eqn:GhostAmp1} can be expanded as
\begin{align}\label{eqn:Exponent}
    &
    \int_{\phi - \theta/2}^{\phi + \theta/2} \ud t\,
    \frac{K(t) \cdot L(t)}{n \cdot k (1 - s)}
    =
    \frac{\theta}{2 n \cdot k s (1 - s)}
    \Big[
        q_{\LCperp}^{2}
        -
        \text{var}(a)
        (e_{l} - s e)^{2}
    \Big] 
    \,,
\end{align}
where the variance is defined by \eqref{Kibblem} and we have changed variables to
\begin{align}\label{eqn:qdef}
    q_{\LCperp}
    =
    l_{\LCperp}
    -
    s
    k_{\LCperp}
    -
    (e_{l} - s e)
    \widehat{a}_{\LCperp}
    \,.
\end{align}
It is a straightforward, though lengthy, exercise to evaluate the complete pre-exponential tensor structure.

To (slightly) reduce the number of terms, we specialize to helicity non-flip, taking $\epsilon = \epsilon^{(+)}$ and $\epsilon^{\prime} = \epsilon^{(-)}$, and defining $\mathcal{A} = \epsilon^{(+)} \cdot \widehat{a}$ and $\mathcal{\bar{A}} = \epsilon^{(-)} \cdot \widehat{a}$. 
One finds
\begin{align}
    \mcT_{\text{ghost}}
    -
    \frac{1}{2}
    \mcT_{\text{gluon}}
    =
    &
    -
    4
    q
    \bar{q}
    -
    \frac{2\,\theta^2}{(1-s) s}
    \bigg(
        1
        -
        \frac{1}{2(1-s)s}
    \bigg)
    (e_{l} - s e)^{2}\,
    \widehat{a}_{\phi}^{2}
    \nonumber\\
    &
    -
    2
    \theta^{2}
    (e_{l} - s e)^{2}
    \bigg[
        -
        \frac{1}{2}
        \mathcal{\bar{A}}_{\phi}
        \mathcal{A}_{\phi}
        +
        2
        \mathcal{\bar{A}}_{\theta}
        \mathcal{A}_{\theta}
        +
        \bigg(
            1
            -
            \frac{2}{s (1-s)}
        \bigg) 
        \mathcal{\bar{A}}_{[\phi}
        \mathcal{A}_{\theta]}
    \bigg]\,,
\end{align}
where $q = (q_{1} + \im q_{2})/\sqrt{2}$ and  $\bar{q} = (q_{1} - \im q_{2})/\sqrt{2}$.
Analytically continuing $\theta \to \theta + \im \delta$, the (Gaussian) integrations over the transverse components $q_{\LCperp}$ can be performed analytically~\cite{Dinu:2013hsd}. 

The scattering amplitude is then
\begin{align}\label{eqn:GhostAmp2}
    \mathcal{M}_{-+}
    =
    &
    \frac{
        \mathrm{g}^{2}
        N\,
        \delta^{\mathsf{a}_{1}\mathsf{a}_{2}}
    }{2 (2\pi)^{2}} 
    \int \ud \phi
    \int_{0}^{\infty} \ud \theta
    \int_{0}^{1} 
    \ud s
    \exp\bigg[
        -
        \im
        \frac{\theta\, \text{var}(a)}{2 n \cdot k\, s (1 - s)}
        \big(e_{l} - s e\big)^{2}
    \bigg] 
    \nonumber\\
    &
    \times
    \bigg\{    
        \frac{2 n \cdot k s (1 - s)}{(\theta + \im \delta)^{2}}
        +
        \frac{\im\theta}{(1 - s) s}
        \bigg(
            1
            -
            \frac{1}{2(1 - s)s}
        \bigg)
        \big(e_{l} - s e\big)^{2}
        \widehat{a}_{\phi}^{2}
        \nonumber\\
        &
        -
        \im
        \theta
        \big(s e - e_{l}\big)^{2}
        \bigg[
            -
            \frac{1}{2}
            \mathcal{\bar{A}}_{\phi}
            \mathcal{A}_{\phi}
            +
            2
            \mathcal{\bar{A}}_{\theta}
            \mathcal{A}_{\theta}
            +
            \bigg(
                1
                -
                \frac{2}{s (1 - s)}
            \bigg) 
            \mathcal{\bar{A}}_{[\phi}
            \mathcal{A}_{\theta]}
        \bigg]
    \bigg\}
    \,.
\end{align}
The first term in this expression is proportional to the UV-divergent free-field scattering amplitude, seen here as a contact divergence $\sim \theta^{-2}$ at $\theta=0$. 
All other terms are UV finite. 
The amplitude is renormalised by subtracting the free-field contribution, following~\cite{Dinu:2014tsa}, after which it may be written
\begin{align}\label{YMnf}
    \mathcal{M}_{-+}
    =
    -
    \im
    &
    \delta^{\mathsf{a}_{1}\mathsf{a}_{2}}
    \frac{
        \mathrm{g}^{2}
        N
    }{8 \pi^{2}} 
    \int_{-\infty}^{\infty} \ud \phi
    \int_{0}^{\infty} \ud \theta \;
    \theta
    \int_{0}^{1} \ud s \;
    \e^{
        -
        \im
        \frac{\theta\, \text{var}(a) (e_{l} - s e)^{2}}{2 n \cdot k\, s (1 - s)}
    }
    \big(e_{l} - s e\big)^{2}
    \nonumber\\
    &
    \times
    \bigg[
        \epsilon^{(+)} \cdot \epsilon^{(-)}
        \bigg(
            \frac{1}{\theta^{2}}
            \partial_{\theta}
            \Big[\theta \text{var}(a)\Big]
            -
            \frac{\widehat{a}_{\phi}^{2}}{(1-s) s}
            \bigg(
                1
                -
                \frac{1}{2(1-s)s}
            \bigg)
        \bigg)
        \nonumber\\
        &
        \qquad
        +
        2
        \cA_{\theta} \bar{\cA}_{\theta}
        -
        \frac{1}{2}
        \cA_{\phi} \bar{\cA}_{\phi}
        +
        \bigg(
            1
            -
            \frac{2}{(1-s)s}
        \bigg) 
        \bar{\cA}_{[\phi}
        \cA_{\theta]}
    \bigg]
    \,.
\end{align}

%% file: main_arxiv_v2.bbl
\providecommand{\href}[2]{#2}\begingroup\raggedright\begin{thebibliography}{10}

\bibitem{Elvang:2013cua}
H.~Elvang and Y.-t. Huang, \emph{{Scattering Amplitudes}},
  \href{http://arxiv.org/abs/1308.1697}{{\tt 1308.1697}}.

\bibitem{Dixon:2013uaa}
L.~J. Dixon, \emph{{A brief introduction to modern amplitude methods}},  in
  \emph{{Theoretical Advanced Study Institute in Elementary Particle Physics:
  Particle Physics: The Higgs Boson and Beyond}}, pp.~31--67, 2014.
\newblock \href{http://arxiv.org/abs/1310.5353}{{\tt 1310.5353}}.
\newblock \href{http://dx.doi.org/10.5170/CERN-2014-008.31}{DOI}.

\bibitem{Cheung:2017pzi}
C.~Cheung, \emph{{TASI Lectures on Scattering Amplitudes}}, pp.~571--623.
\newblock 2018.
\newblock \href{http://arxiv.org/abs/1708.03872}{{\tt 1708.03872}}.

\bibitem{Furry:1951zz}
W.~H. Furry, \emph{{On Bound States and Scattering in Positron Theory}},
  \href{http://dx.doi.org/10.1103/PhysRev.81.915}{\emph{Phys. Rev.} {\bf 81}
  (1951) 115--124}.

\bibitem{DeWitt:1967ub}
B.~S. DeWitt, \emph{{Quantum Theory of Gravity. 2. The Manifestly Covariant
  Theory}}, \href{http://dx.doi.org/10.1103/PhysRev.162.1195}{\emph{Phys. Rev.}
  {\bf 162} (1967) 1195--1239}.

\bibitem{tHooft:1975uxh}
G.~'t~Hooft, \emph{{The Background Field Method in Gauge Field Theories}},  in
  \emph{{12th Annual Winter School of Theoretical Physics}}, pp.~345--369,
  1975.

\bibitem{Boulware:1980av}
D.~G. Boulware, \emph{{Gauge Dependence of the Effective Action}},
  \href{http://dx.doi.org/10.1103/PhysRevD.23.389}{\emph{Phys. Rev. D} {\bf 23}
  (1981) 389}.

\bibitem{Abbott:1981ke}
L.~F. Abbott, \emph{{Introduction to the Background Field Method}}, {\emph{Acta
  Phys. Polon. B} {\bf 13} (1982) 33}.

\bibitem{Nikishov:1964zza}
A.~I. Nikishov and V.~I. Ritus, \emph{{Quantum Processes in the Field of a
  Plane Electromagnetic Wave and in a Constant Field 1}}, {\emph{Sov. Phys.
  JETP} {\bf 19} (1964) 529--541}.

\bibitem{Ritus:1985}
V.~I. Ritus, \emph{Quantum effects of the interaction of elementary particles
  with an intense electromagnetic field},
  \href{http://dx.doi.org/10.1007/BF01120220}{\emph{Journal of Soviet Laser
  Research} {\bf 6} (Sep, 1985) 497--617}.

\bibitem{DiPiazza:2011tq}
A.~Di~Piazza, C.~Muller, K.~Z. Hatsagortsyan and C.~H. Keitel, \emph{{Extremely
  high-intensity laser interactions with fundamental quantum systems}},
  \href{http://dx.doi.org/10.1103/RevModPhys.84.1177}{\emph{Rev. Mod. Phys.}
  {\bf 84} (2012) 1177}, [\href{http://arxiv.org/abs/1111.3886}{{\tt
  1111.3886}}].

\bibitem{Seipt:2017ckc}
D.~Seipt, \emph{{Volkov States and Non-linear Compton Scattering in Short and
  Intense Laser Pulses}},  in \emph{{Quantum Field Theory at the Limits}: {from
  Strong Fields to Heavy Quarks}}, pp.~24--43, 2017.
\newblock \href{http://arxiv.org/abs/1701.03692}{{\tt 1701.03692}}.
\newblock \href{http://dx.doi.org/10.3204/DESY-PROC-2016-04/Seipt}{DOI}.

\bibitem{King:2015tba}
B.~King and T.~Heinzl, \emph{{Measuring Vacuum Polarisation with High Power
  Lasers}},  \href{http://arxiv.org/abs/1510.08456}{{\tt 1510.08456}}.

\bibitem{Bamber:1999zt}
C.~Bamber et~al., \emph{{Studies of nonlinear QED in collisions of 46.6-GeV
  electrons with intense laser pulses}},
  \href{http://dx.doi.org/10.1103/PhysRevD.60.092004}{\emph{Phys. Rev.} {\bf
  D60} (1999) 092004}.

\bibitem{Abramowicz:2021zja}
H.~Abramowicz et~al., \emph{{Conceptual Design Report for the LUXE
  Experiment}},  \href{http://arxiv.org/abs/2102.02032}{{\tt 2102.02032}}.

\bibitem{E320}
S.~Meuren, \emph{{Probing Strong-field QED at FACET-II (SLAC E-320)}},
  {\emph{{Talk presented at FACET-II Science Workshop}} (2019) }.

\bibitem{Toll:1952rq}
J.~S. Toll, \emph{{The dispersion relation for light and its application to
  problems involving electron pairs}}.
\newblock PhD thesis, Princeton U., 1952.

\bibitem{Heinzl:2006xc}
T.~Heinzl, B.~Liesfeld, K.-U. Amthor, H.~Schwoerer, R.~Sauerbrey and A.~Wipf,
  \emph{{On the observation of vacuum birefringence}},
  \href{http://dx.doi.org/10.1016/j.optcom.2006.06.053}{\emph{Opt. Commun.}
  {\bf 267} (2006) 318--321}, [\href{http://arxiv.org/abs/hep-ph/0601076}{{\tt
  hep-ph/0601076}}].

\bibitem{Dunne:2004nc}
G.~V. Dunne, \emph{{Heisenberg-Euler effective Lagrangians: Basics and
  extensions}},  in \emph{{From fields to strings: Circumnavigating theoretical
  physics. Ian Kogan memorial collection (3 volume set)}} (M.~Shifman,
  A.~Vainshtein and J.~Wheater, eds.), pp.~445--522.
\newblock 6, 2004.
\newblock \href{http://arxiv.org/abs/hep-th/0406216}{{\tt hep-th/0406216}}.
\newblock \href{http://dx.doi.org/10.1142/9789812775344_0014}{DOI}.

\bibitem{Gies2017}
H.~Gies and F.~Karbstein, \emph{{An Addendum to the Heisenberg-Euler effective
  action beyond one loop}},
  \href{http://dx.doi.org/10.1007/JHEP03(2017)108}{\emph{JHEP} {\bf 03} (2016)
  108}, [\href{http://arxiv.org/abs/1612.07251}{{\tt 1612.07251}}].

\bibitem{Dinu:2013gaa}
V.~Dinu, T.~Heinzl, A.~Ilderton, M.~Marklund and G.~Torgrimsson, \emph{{Vacuum
  refractive indices and helicity flip in strong-field QED}},
  \href{http://dx.doi.org/10.1103/PhysRevD.89.125003}{\emph{Phys. Rev. D} {\bf
  89} (2014) 125003}, [\href{http://arxiv.org/abs/1312.6419}{{\tt 1312.6419}}].

\bibitem{Adamo:2017nia}
T.~Adamo, E.~Casali, L.~Mason and S.~Nekovar, \emph{{Scattering on plane waves
  and the double copy}},
  \href{http://dx.doi.org/10.1088/1361-6382/aa9961}{\emph{Class. Quant. Grav.}
  {\bf 35} (2018) 015004}, [\href{http://arxiv.org/abs/1706.08925}{{\tt
  1706.08925}}].

\bibitem{Farrow:2018yni}
J.~A. Farrow, A.~E. Lipstein and P.~McFadden, \emph{{Double copy structure of
  CFT correlators}},
  \href{http://dx.doi.org/10.1007/JHEP02(2019)130}{\emph{JHEP} {\bf 02} (2019)
  130}, [\href{http://arxiv.org/abs/1812.11129}{{\tt 1812.11129}}].

\bibitem{Adamo:2018mpq}
T.~Adamo, E.~Casali, L.~Mason and S.~Nekovar, \emph{{Plane wave backgrounds and
  colour-kinematics duality}},
  \href{http://dx.doi.org/10.1007/JHEP02(2019)198}{\emph{JHEP} {\bf 02} (2019)
  198}, [\href{http://arxiv.org/abs/1810.05115}{{\tt 1810.05115}}].

\bibitem{Lipstein:2019mpu}
A.~E. Lipstein and P.~McFadden, \emph{{Double copy structure and the flat space
  limit of conformal correlators in even dimensions}},
  \href{http://dx.doi.org/10.1103/PhysRevD.101.125006}{\emph{Phys. Rev. D} {\bf
  101} (2020) 125006}, [\href{http://arxiv.org/abs/1912.10046}{{\tt
  1912.10046}}].

\bibitem{Adamo:2020qru}
T.~Adamo and A.~Ilderton, \emph{{Classical and quantum double copy of
  back-reaction}}, \href{http://dx.doi.org/10.1007/JHEP09(2020)200}{\emph{JHEP}
  {\bf 09} (2020) 200}, [\href{http://arxiv.org/abs/2005.05807}{{\tt
  2005.05807}}].

\bibitem{Albayrak:2020fyp}
S.~Albayrak, S.~Kharel and D.~Meltzer, \emph{{On duality of color and
  kinematics in (A)dS momentum space}},
  \href{http://arxiv.org/abs/2012.10460}{{\tt 2012.10460}}.

\bibitem{Armstrong:2020woi}
C.~Armstrong, A.~E. Lipstein and J.~Mei, \emph{{Color/kinematics duality in
  AdS$_{4}$}}, \href{http://dx.doi.org/10.1007/JHEP02(2021)194}{\emph{JHEP}
  {\bf 02} (2021) 194}, [\href{http://arxiv.org/abs/2012.02059}{{\tt
  2012.02059}}].

\bibitem{Casali:2020vuy}
E.~Casali and A.~Puhm, \emph{{Double Copy for Celestial Amplitudes}},
  \href{http://dx.doi.org/10.1103/PhysRevLett.126.101602}{\emph{Phys. Rev.
  Lett.} {\bf 126} (2021) 101602}, [\href{http://arxiv.org/abs/2007.15027}{{\tt
  2007.15027}}].

\bibitem{Casali:2020uvr}
E.~Casali and A.~Sharma, \emph{{Celestial double copy from the worldsheet}},
  \href{http://arxiv.org/abs/2011.10052}{{\tt 2011.10052}}.

\bibitem{Pasterski:2020pdk}
S.~Pasterski and A.~Puhm, \emph{{Shifting Spin on the Celestial Sphere}},
  \href{http://arxiv.org/abs/2012.15694}{{\tt 2012.15694}}.

\bibitem{Dunne:2002qf}
G.~V. Dunne and C.~Schubert, \emph{{Two loop selfdual Euler-Heisenberg
  Lagrangians. 1. Real part and helicity amplitudes}},
  \href{http://dx.doi.org/10.1088/1126-6708/2002/08/053}{\emph{JHEP} {\bf 08}
  (2002) 053}, [\href{http://arxiv.org/abs/hep-th/0205004}{{\tt
  hep-th/0205004}}].

\bibitem{Dunne:2002qg}
G.~V. Dunne and C.~Schubert, \emph{{Two loop selfdual Euler-Heisenberg
  Lagrangians. 2. Imaginary part and Borel analysis}},
  \href{http://dx.doi.org/10.1088/1126-6708/2002/06/042}{\emph{JHEP} {\bf 06}
  (2002) 042}, [\href{http://arxiv.org/abs/hep-th/0205005}{{\tt
  hep-th/0205005}}].

\bibitem{Adamo:2020syc}
T.~Adamo, L.~Mason and A.~Sharma, \emph{{MHV scattering of gluons and gravitons
  in chiral strong fields}},
  \href{http://dx.doi.org/10.1103/PhysRevLett.125.041602}{\emph{Phys. Rev.
  Lett.} {\bf 125} (2020) 041602}, [\href{http://arxiv.org/abs/2003.13501}{{\tt
  2003.13501}}].

\bibitem{Adamo:2020yzi}
T.~Adamo, L.~Mason and A.~Sharma, \emph{{Gluon scattering on self-dual
  radiative gauge fields}},  \href{http://arxiv.org/abs/2010.14996}{{\tt
  2010.14996}}.

\bibitem{Mahlon:1993fe}
G.~Mahlon, \emph{{One loop multi - photon helicity amplitudes}},
  \href{http://dx.doi.org/10.1103/PhysRevD.49.2197}{\emph{Phys. Rev. D} {\bf
  49} (1994) 2197--2210}, [\href{http://arxiv.org/abs/hep-ph/9311213}{{\tt
  hep-ph/9311213}}].

\bibitem{Bern:1993qk}
Z.~Bern, G.~Chalmers, L.~J. Dixon and D.~A. Kosower, \emph{{One loop N gluon
  amplitudes with maximal helicity violation via collinear limits}},
  \href{http://dx.doi.org/10.1103/PhysRevLett.72.2134}{\emph{Phys. Rev. Lett.}
  {\bf 72} (1994) 2134--2137}, [\href{http://arxiv.org/abs/hep-ph/9312333}{{\tt
  hep-ph/9312333}}].

\bibitem{Mahlon:1993si}
G.~Mahlon, \emph{{Multi - gluon helicity amplitudes involving a quark loop}},
  \href{http://dx.doi.org/10.1103/PhysRevD.49.4438}{\emph{Phys. Rev. D} {\bf
  49} (1994) 4438--4453}, [\href{http://arxiv.org/abs/hep-ph/9312276}{{\tt
  hep-ph/9312276}}].

\bibitem{Bern:1994ju}
Z.~Bern, L.~J. Dixon, D.~C. Dunbar and D.~A. Kosower, \emph{{One loop gauge
  theory amplitudes with an arbitrary number of external legs}},  in
  \emph{{Workshop on Continuous Advances in QCD}}, pp.~3--21, 2, 1994.
\newblock \href{http://arxiv.org/abs/hep-ph/9405248}{{\tt hep-ph/9405248}}.

\bibitem{Bern:2005ji}
Z.~Bern, L.~J. Dixon and D.~A. Kosower, \emph{{The last of the finite loop
  amplitudes in QCD}},
  \href{http://dx.doi.org/10.1103/PhysRevD.72.125003}{\emph{Phys. Rev. D} {\bf
  72} (2005) 125003}, [\href{http://arxiv.org/abs/hep-ph/0505055}{{\tt
  hep-ph/0505055}}].

\bibitem{Bern:1996ja}
Z.~Bern, L.~J. Dixon, D.~C. Dunbar and D.~A. Kosower, \emph{{One loop selfdual
  and N=4 superYang-Mills}},
  \href{http://dx.doi.org/10.1016/S0370-2693(96)01676-0}{\emph{Phys. Lett. B}
  {\bf 394} (1997) 105--115}, [\href{http://arxiv.org/abs/hep-th/9611127}{{\tt
  hep-th/9611127}}].

\bibitem{Britto:2020crg}
R.~Britto, G.~R. Jehu and A.~Orta, \emph{{The dimension-shift conjecture for
  one-loop amplitudes}},  \href{http://arxiv.org/abs/2011.13821}{{\tt
  2011.13821}}.

\bibitem{Bern:2015xsa}
Z.~Bern, C.~Cheung, H.-H. Chi, S.~Davies, L.~Dixon and J.~Nohle,
  \emph{{Evanescent Effects Can Alter Ultraviolet Divergences in Quantum
  Gravity without Physical Consequences}},
  \href{http://dx.doi.org/10.1103/PhysRevLett.115.211301}{\emph{Phys. Rev.
  Lett.} {\bf 115} (2015) 211301}, [\href{http://arxiv.org/abs/1507.06118}{{\tt
  1507.06118}}].

\bibitem{Bern:2017puu}
Z.~Bern, H.-H. Chi, L.~Dixon and A.~Edison, \emph{{Two-Loop Renormalization of
  Quantum Gravity Simplified}},
  \href{http://dx.doi.org/10.1103/PhysRevD.95.046013}{\emph{Phys. Rev. D} {\bf
  95} (2017) 046013}, [\href{http://arxiv.org/abs/1701.02422}{{\tt
  1701.02422}}].

\bibitem{Chattopadhyay:2020oxe}
P.~Chattopadhyay and K.~Krasnov, \emph{{One-loop same helicity four-point
  amplitude from shifts}},
  \href{http://dx.doi.org/10.1007/JHEP06(2020)082}{\emph{JHEP} {\bf 06} (2020)
  082}, [\href{http://arxiv.org/abs/2002.11390}{{\tt 2002.11390}}].

\bibitem{Grisaru:1976vm}
M.~T. Grisaru, H.~N. Pendleton and P.~van Nieuwenhuizen, \emph{{Supergravity
  and the S Matrix}},
  \href{http://dx.doi.org/10.1103/PhysRevD.15.996}{\emph{Phys. Rev. D} {\bf 15}
  (1977) 996}.

\bibitem{Grisaru:1977px}
M.~T. Grisaru and H.~N. Pendleton, \emph{{Some Properties of Scattering
  Amplitudes in Supersymmetric Theories}},
  \href{http://dx.doi.org/10.1016/0550-3213(77)90277-2}{\emph{Nucl. Phys. B}
  {\bf 124} (1977) 81--92}.

\bibitem{Parke:1985pn}
S.~J. Parke and T.~R. Taylor, \emph{{Perturbative QCD Utilizing Extended
  Supersymmetry}},
  \href{http://dx.doi.org/10.1016/0370-2693(85)91216-X}{\emph{Phys. Lett. B}
  {\bf 157} (1985) 81}.

\bibitem{Kunszt:1985mg}
Z.~Kunszt, \emph{{Combined Use of the Calkul Method and N=1 Supersymmetry to
  Calculate QCD Six Parton Processes}},
  \href{http://dx.doi.org/10.1016/0550-3213(86)90319-6}{\emph{Nucl. Phys. B}
  {\bf 271} (1986) 333--348}.

\bibitem{Brinkmann:1925fr}
H.~Brinkmann, \emph{{Einstein spaces which are mapped conformally on each
  other}}, \href{http://dx.doi.org/10.1007/BF01208647}{\emph{Math. Ann.} {\bf
  94} (1925) 119--145}.

\bibitem{Ilderton:2018lsf}
A.~Ilderton, \emph{{Screw-symmetric gravitational waves: a double copy of the
  vortex}}, \href{http://dx.doi.org/10.1016/j.physletb.2018.04.069}{\emph{Phys.
  Lett. B} {\bf 782} (2018) 22--27},
  [\href{http://arxiv.org/abs/1804.07290}{{\tt 1804.07290}}].

\bibitem{Schwinger:1951nm}
J.~S. Schwinger, \emph{{On gauge invariance and vacuum polarization}},
  \href{http://dx.doi.org/10.1103/PhysRev.82.664}{\emph{Phys. Rev.} {\bf 82}
  (1951) 664--679}.

\bibitem{Kibble:1975vz}
T.~W.~B. Kibble, A.~Salam and J.~A. Strathdee, \emph{{Intensity Dependent Mass
  Shift and Symmetry Breaking}},
  \href{http://dx.doi.org/10.1016/0550-3213(75)90581-7}{\emph{Nucl. Phys. B}
  {\bf 96} (1975) 255--263}.

\bibitem{Harvey:2012ie}
C.~Harvey, T.~Heinzl, A.~Ilderton and M.~Marklund, \emph{{Intensity-Dependent
  Electron Mass Shift in a Laser Field: Existence, Universality, and
  Detection}},
  \href{http://dx.doi.org/10.1103/PhysRevLett.109.100402}{\emph{Phys. Rev.
  Lett.} {\bf 109} (2012) 100402}, [\href{http://arxiv.org/abs/1203.6077}{{\tt
  1203.6077}}].

\bibitem{Adamo:2019zmk}
T.~Adamo and A.~Ilderton, \emph{{Gluon helicity flip in a plane wave
  background}}, \href{http://dx.doi.org/10.1007/JHEP06(2019)015}{\emph{JHEP}
  {\bf 06} (2019) 015}, [\href{http://arxiv.org/abs/1903.01491}{{\tt
  1903.01491}}].

\bibitem{Shore:2007um}
G.~M. Shore, \emph{{Superluminality and UV completion}},
  \href{http://dx.doi.org/10.1016/j.nuclphysb.2007.03.034}{\emph{Nucl. Phys. B}
  {\bf 778} (2007) 219--258}, [\href{http://arxiv.org/abs/hep-th/0701185}{{\tt
  hep-th/0701185}}].

\bibitem{Heinzl:2010vg}
T.~Heinzl, A.~Ilderton and M.~Marklund, \emph{{Finite size effects in
  stimulated laser pair production}},
  \href{http://dx.doi.org/10.1016/j.physletb.2010.07.044}{\emph{Phys. Lett. B}
  {\bf 692} (2010) 250--256}, [\href{http://arxiv.org/abs/1002.4018}{{\tt
  1002.4018}}].

\bibitem{Meuren:2013oya}
S.~Meuren, C.~H. Keitel and A.~Di~Piazza, \emph{{Polarization operator for
  plane-wave background fields}},
  \href{http://dx.doi.org/10.1103/PhysRevD.88.013007}{\emph{Phys. Rev. D} {\bf
  88} (2013) 013007}, [\href{http://arxiv.org/abs/1304.7672}{{\tt 1304.7672}}].

\bibitem{KibbleShift}
T.~W.~B. Kibble, \emph{Frequency shift in high-intensity compton scattering},
  \href{http://dx.doi.org/10.1103/PhysRev.138.B740}{\emph{Phys. Rev.} {\bf 138}
  (May, 1965) B740--B753}.

\bibitem{Frantz}
L.~M. Frantz, \emph{Compton scattering of an intense photon beam},
  \href{http://dx.doi.org/10.1103/PhysRev.139.B1326}{\emph{Phys. Rev.} {\bf
  139} (Sep, 1965) B1326--B1336}.

\bibitem{Gavrilov:1990qa}
S.~P. Gavrilov and D.~M. Gitman, \emph{{QED interpretation of external field
  and external current. (In Russian)}}, {\emph{Sov. J. Nucl. Phys.} {\bf 51}
  (1990) 1040--1045}.

\bibitem{Ilderton:2017xbj}
A.~Ilderton and D.~Seipt, \emph{{Backreaction on background fields: A coherent
  state approach}},
  \href{http://dx.doi.org/10.1103/PhysRevD.97.016007}{\emph{Phys. Rev. D} {\bf
  97} (2018) 016007}, [\href{http://arxiv.org/abs/1709.10085}{{\tt
  1709.10085}}].

\bibitem{Seipt:2010ya}
D.~Seipt and B.~Kampfer, \emph{{Non-Linear Compton Scattering of Ultrashort and
  Ultraintense Laser Pulses}},
  \href{http://dx.doi.org/10.1103/PhysRevA.83.022101}{\emph{Phys. Rev. A} {\bf
  83} (2011) 022101}, [\href{http://arxiv.org/abs/1010.3301}{{\tt 1010.3301}}].

\bibitem{Ilderton:2019bop}
A.~Ilderton, B.~King and A.~J. Macleod, \emph{{Absorption cross section in an
  intense plane wave background}},
  \href{http://dx.doi.org/10.1103/PhysRevD.100.076002}{\emph{Phys. Rev. D} {\bf
  100} (2019) 076002}, [\href{http://arxiv.org/abs/1907.12835}{{\tt
  1907.12835}}].

\bibitem{Bardeen:1995gk}
W.~A. Bardeen, \emph{{Selfdual Yang-Mills theory, integrability and multiparton
  amplitudes}}, \href{http://dx.doi.org/10.1143/PTPS.123.1}{\emph{Prog. Theor.
  Phys. Suppl.} {\bf 123} (1996) 1--8}.

\bibitem{Cangemi:1996rx}
D.~Cangemi, \emph{{Selfdual Yang-Mills theory and one loop like - helicity QCD
  multi - gluon amplitudes}},
  \href{http://dx.doi.org/10.1016/S0550-3213(96)00586-X}{\emph{Nucl. Phys. B}
  {\bf 484} (1997) 521--537}, [\href{http://arxiv.org/abs/hep-th/9605208}{{\tt
  hep-th/9605208}}].

\bibitem{Bernicot:2008th}
C.~Bernicot, \emph{{Light-light amplitude from generalized unitarity in massive
  QED}},  \href{http://arxiv.org/abs/0804.0749}{{\tt 0804.0749}}.

\bibitem{Gastmans:1990xh}
R.~Gastmans and T.~Wu, \emph{{The Ubiquitous photon: Helicity method for QED
  and QCD}}, vol.~80.
\newblock Clarendon, 1990.

\bibitem{Gaunt:2011xd}
J.~R. Gaunt and W.~J. Stirling, \emph{{Double Parton Scattering Singularity in
  One-Loop Integrals}},
  \href{http://dx.doi.org/10.1007/JHEP06(2011)048}{\emph{JHEP} {\bf 06} (2011)
  048}, [\href{http://arxiv.org/abs/1103.1888}{{\tt 1103.1888}}].

\bibitem{Dixon:2016epj}
L.~J. Dixon and I.~Esterlis, \emph{{All orders results for self-crossing Wilson
  loops mimicking double parton scattering}},
  \href{http://dx.doi.org/10.1007/JHEP07(2016)116}{\emph{JHEP} {\bf 07} (2016)
  116}, [\href{http://arxiv.org/abs/1602.02107}{{\tt 1602.02107}}].

\bibitem{Stieberger:2015kia}
S.~Stieberger and T.~R. Taylor, \emph{{Subleading terms in the collinear limit
  of Yang\textendash{}Mills amplitudes}},
  \href{http://dx.doi.org/10.1016/j.physletb.2015.09.075}{\emph{Phys. Lett. B}
  {\bf 750} (2015) 587--590}, [\href{http://arxiv.org/abs/1508.01116}{{\tt
  1508.01116}}].

\bibitem{Nandan:2016ohb}
D.~Nandan, J.~Plefka and W.~Wormsbecher, \emph{{Collinear limits beyond the
  leading order from the scattering equations}},
  \href{http://dx.doi.org/10.1007/JHEP02(2017)038}{\emph{JHEP} {\bf 02} (2017)
  038}, [\href{http://arxiv.org/abs/1608.04730}{{\tt 1608.04730}}].

\bibitem{Bhattacharya:2018vph}
A.~Bhattacharya, I.~Moult, I.~W. Stewart and G.~Vita, \emph{{Helicity Methods
  for High Multiplicity Subleading Soft and Collinear Limits}},
  \href{http://dx.doi.org/10.1007/JHEP05(2019)192}{\emph{JHEP} {\bf 05} (2019)
  192}, [\href{http://arxiv.org/abs/1812.06950}{{\tt 1812.06950}}].

\bibitem{Mangano:1990by}
M.~L. Mangano and S.~J. Parke, \emph{{Multiparton amplitudes in gauge
  theories}}, \href{http://dx.doi.org/10.1016/0370-1573(91)90091-Y}{\emph{Phys.
  Rept.} {\bf 200} (1991) 301--367},
  [\href{http://arxiv.org/abs/hep-th/0509223}{{\tt hep-th/0509223}}].

\bibitem{Bern:1990ux}
Z.~Bern and D.~A. Kosower, \emph{{Color decomposition of one loop amplitudes in
  gauge theories}},
  \href{http://dx.doi.org/10.1016/0550-3213(91)90567-H}{\emph{Nucl. Phys. B}
  {\bf 362} (1991) 389--448}.

\bibitem{Bern:1991aq}
Z.~Bern and D.~A. Kosower, \emph{{The Computation of loop amplitudes in gauge
  theories}}, \href{http://dx.doi.org/10.1016/0550-3213(92)90134-W}{\emph{Nucl.
  Phys. B} {\bf 379} (1992) 451--561}.

\bibitem{Bern:1995db}
Z.~Bern and A.~G. Morgan, \emph{{Massive loop amplitudes from unitarity}},
  \href{http://dx.doi.org/10.1016/0550-3213(96)00078-8}{\emph{Nucl. Phys. B}
  {\bf 467} (1996) 479--509}, [\href{http://arxiv.org/abs/hep-ph/9511336}{{\tt
  hep-ph/9511336}}].

\bibitem{Trautman:1980bj}
A.~Trautman, \emph{{A class of null solutions to Yang-Mills equations}},
  \href{http://dx.doi.org/10.1088/0305-4470/13/1/001}{\emph{J. Phys. A} {\bf
  13} (1980) L1--L4}.

\bibitem{Sangal:2021qeg}
M.~Sangal, C.~H. Keitel and M.~Tamburini, \emph{{Observing Light-by-Light
  Scattering in Vacuum with an Asymmetric Photon Collider}},
  \href{http://arxiv.org/abs/2101.02671}{{\tt 2101.02671}}.

\bibitem{Gies:2021ymf}
H.~Gies, F.~Karbstein and L.~Klar, \emph{{Quantum vacuum signatures in
  multi-color laser pulse collisions}},
  \href{http://arxiv.org/abs/2101.04461}{{\tt 2101.04461}}.

\bibitem{Duff:1979bk}
M.~J. Duff and C.~J. Isham, \emph{{Selfduality, Helicity, and Supersymmetry:
  The Scattering of Light by Light}},
  \href{http://dx.doi.org/10.1016/0370-2693(79)90807-4}{\emph{Phys. Lett. B}
  {\bf 86} (1979) 157--160}.

\bibitem{Rebhan:2017zdx}
A.~Rebhan and G.~Turk, \emph{{Polarization effects in light-by-light
  scattering: Euler\textendash{}Heisenberg versus Born\textendash{}Infeld}},
  \href{http://dx.doi.org/10.1142/S0217751X17500531}{\emph{Int. J. Mod. Phys.
  A} {\bf 32} (2017) 1750053}, [\href{http://arxiv.org/abs/1701.07375}{{\tt
  1701.07375}}].

\bibitem{Cheung:2018oki}
C.~Cheung, K.~Kampf, J.~Novotny, C.-H. Shen, J.~Trnka and C.~Wen, \emph{{Vector
  Effective Field Theories from Soft Limits}},
  \href{http://dx.doi.org/10.1103/PhysRevLett.120.261602}{\emph{Phys. Rev.
  Lett.} {\bf 120} (2018) 261602}, [\href{http://arxiv.org/abs/1801.01496}{{\tt
  1801.01496}}].

\bibitem{Ilderton:2016qpj}
A.~Ilderton and G.~Torgrimsson, \emph{{Worldline approach to helicity flip in
  plane waves}},
  \href{http://dx.doi.org/10.1103/PhysRevD.93.085006}{\emph{Phys. Rev. D} {\bf
  93} (2016) 085006}, [\href{http://arxiv.org/abs/1601.05021}{{\tt
  1601.05021}}].

\bibitem{Gounaris:2005ey}
G.~J. Gounaris and F.~M. Renard, \emph{{Helicity conservation in gauge boson
  scattering at high energy}},
  \href{http://dx.doi.org/10.1103/PhysRevLett.94.131601}{\emph{Phys. Rev.
  Lett.} {\bf 94} (2005) 131601},
  [\href{http://arxiv.org/abs/hep-ph/0501046}{{\tt hep-ph/0501046}}].

\bibitem{Gounaris:2006zm}
G.~J. Gounaris and F.~M. Renard, \emph{{Addendum to `Helicity conservation in
  gauge boson scattering at high energy'}},
  \href{http://dx.doi.org/10.1103/PhysRevD.73.097301}{\emph{Phys. Rev. D} {\bf
  73} (2006) 097301}, [\href{http://arxiv.org/abs/hep-ph/0604041}{{\tt
  hep-ph/0604041}}].

\bibitem{Ilderton:2020gno}
A.~Ilderton, B.~King and S.~Tang, \emph{{Loop spin effects in intense
  background fields}},
  \href{http://dx.doi.org/10.1103/PhysRevD.102.076013}{\emph{Phys. Rev. D} {\bf
  102} (2020) 076013}, [\href{http://arxiv.org/abs/2008.08578}{{\tt
  2008.08578}}].

\bibitem{Bern:2019prr}
Z.~Bern, J.~J. Carrasco, M.~Chiodaroli, H.~Johansson and R.~Roiban, \emph{{The
  Duality Between Color and Kinematics and its Applications}},
  \href{http://arxiv.org/abs/1909.01358}{{\tt 1909.01358}}.

\bibitem{Adamo:2017sze}
T.~Adamo, E.~Casali, L.~Mason and S.~Nekovar, \emph{{Amplitudes on plane waves
  from ambitwistor strings}},
  \href{http://dx.doi.org/10.1007/JHEP11(2017)160}{\emph{JHEP} {\bf 11} (2017)
  160}, [\href{http://arxiv.org/abs/1708.09249}{{\tt 1708.09249}}].

\bibitem{DiPiazza:2007yx}
A.~Di~Piazza, A.~I. Milstein and C.~H. Keitel, \emph{{Photon splitting in a
  laser field}},
  \href{http://dx.doi.org/10.1103/PhysRevA.76.032103}{\emph{Phys. Rev. A} {\bf
  76} (2007) 032103}, [\href{http://arxiv.org/abs/0704.0695}{{\tt 0704.0695}}].

\bibitem{Nagy:2006xy}
Z.~Nagy and D.~E. Soper, \emph{{Numerical integration of one-loop Feynman
  diagrams for N-photon amplitudes}},
  \href{http://dx.doi.org/10.1103/PhysRevD.74.093006}{\emph{Phys. Rev. D} {\bf
  74} (2006) 093006}, [\href{http://arxiv.org/abs/hep-ph/0610028}{{\tt
  hep-ph/0610028}}].

\bibitem{Ossola:2007bb}
G.~Ossola, C.~G. Papadopoulos and R.~Pittau, \emph{{Numerical evaluation of
  six-photon amplitudes}},
  \href{http://dx.doi.org/10.1088/1126-6708/2007/07/085}{\emph{JHEP} {\bf 07}
  (2007) 085}, [\href{http://arxiv.org/abs/0704.1271}{{\tt 0704.1271}}].

\bibitem{Binoth:2007ca}
T.~Binoth, G.~Heinrich, T.~Gehrmann and P.~Mastrolia, \emph{{Six-Photon
  Amplitudes}},
  \href{http://dx.doi.org/10.1016/j.physletb.2007.04.032}{\emph{Phys. Lett. B}
  {\bf 649} (2007) 422--426}, [\href{http://arxiv.org/abs/hep-ph/0703311}{{\tt
  hep-ph/0703311}}].

\bibitem{Casher:1976ae}
A.~Casher, \emph{{Gauge Fields on the Null Plane}},
  \href{http://dx.doi.org/10.1103/PhysRevD.14.452}{\emph{Phys. Rev. D} {\bf 14}
  (1976) 452}.

\bibitem{Dinu:2013hsd}
V.~Dinu, \emph{{Exact final state integrals for strong field QED}},
  \href{http://dx.doi.org/10.1103/PhysRevA.87.052101}{\emph{Phys. Rev. A} {\bf
  87} (2013) 052101}, [\href{http://arxiv.org/abs/1302.1513}{{\tt 1302.1513}}].

\bibitem{Dinu:2014tsa}
V.~Dinu, T.~Heinzl, A.~Ilderton, M.~Marklund and G.~Torgrimsson, \emph{{Photon
  polarization in light-by-light scattering: Finite size effects}},
  \href{http://dx.doi.org/10.1103/PhysRevD.90.045025}{\emph{Phys. Rev. D} {\bf
  90} (2014) 045025}, [\href{http://arxiv.org/abs/1405.7291}{{\tt 1405.7291}}].

\end{thebibliography}\endgroup
